# Structured free-space optical fields for transverse and longitudinal control of electron matter waves

Sven Ebel[1,*] and Nahid Talebi[2,3,⊥]

[1]POLIMA—Center for Polariton-driven Light-Matter Interactions, University of Southern Denmark, 5230 Odense, Denmark

[2]Institute of Experimental and Applied Physics, Kiel University, 24098 Kiel, Germany

[3]Kiel Nano, Surface and Interface Science KiNSIS, Kiel University, 24118 Kiel, Germany

⊥ talebi@physik.uni-kiel.de, * sleb@mci.sdu.dk

**Abstract**

Controlling free-electron momentum states is of high interest in electron microscopy to achieve momentum and energy resolved probing and manipulation of physical systems. Free-electron and light interactions have emerged as a powerful technique to accomplish this. Here, we demonstrate both longitudinal and transverse phase control of a slow electron wavepacket by extending the Kapitza-Dirac effect to spatially-structured pulsed laser beams. This extension enables both inelastic and elastic stimulated Compton scattering. The interaction reveals the formation of distinct electron transverse momentum orders, each demonstrating a comb-like electron energy spectrum. By exerting complete control over light parameters, including wavelength, field intensity, pulse duration, and spatial mode order, as well as their combinations, it is possible to coherently control the population of these electron energy-momentum states that are separated by a few meV energy and multiple photon momentum orders. This free-space electron-light interaction phenomenon possesses the capability to coherently control the energy and momentum of electron beams in electron microscopes. Moreover, it has the potential to facilitate the selective probing of various material excitations, including plasmons, excitons, and phonons, and performing Talbot-Lau matter-wave interferometry with transversely shaped electron beams.

**Introduction**

Manipulation of the electron beam plays a vital role in electron microscopes. The very first electron microscope pioneered the manipulation of electron beams by controlling the trajectory of the electron beam with electron lenses applying tailored electro- and magneto static fields[1]. More recent developments based on hologram masks[2–4] and electron phase plates[5–7] have demonstrated that spatial electron beam manipulation is possible with techniques other than electron lenses. These methods, known as transverse phase manipulation techniques, are based on the principle of introducing a spatial phase shift to the electron wave, allowing the preparation of an electron beam with an arbitrary amplitude structure after propagation in vacuum. With the advent of electron-wave manipulation through interaction with light, the approach of manipulating the phase of the electron waves has been further generalized to the separate manipulation of the transverse and longitudinal phase of the electron beam, translating to a momentum and energy manipulation. One such technique is photon-induced near-field electron microscopy (PINEM). PINEM has enabled the manipulation of both longitudinal and transverse electron phase[8–12]. The longitudinal phase manipulation results in the formation of a comb like electron energy spectrum and has enabled the full control of transverse and longitudinal phase of the electron wavepacket, even reaching simultaneous control of both transverse and longitudinal phase[13,14]. Free electron light interactions have also been applied to construct ponderomotive phase plates made from laser light to enable electron-beam amplitude modulation[6,7,15,16]. More recently the imprinting of the spatial light structure onto the electron beam was demonstrated in free space[7,17,18].

Electron light interactions are usually difficult to achieve since the net absorption or emission of a photon by an electron in free-space is forbidden due to energy-momentum conservation. Therefore, overcoming the mismatch of electron and light dispersion relations is the fundamental principle of any electron-light interaction mechanism. One approach to achieving this is by using a material with the refractive index larger than $c/v_{\rm el}$. An electron traveling through such a material emits Cherenkov radiation. PINEM is enabled through the introduction of evanescent near-field light by nanostructures, facilitating the fulfillment of energy and momentum requirements for electron-photon interactions. This longitudinal phase modulation is linked to the quantized absorption and emission of photons, defining the subsequent manipulation of the electron energy spectrum.

At the absence of materials, electron-light interaction cross-section is generally negligible. The elastic scattering (the transfer of momentum) of an electron is possible for higher-order photon processes (e.g. multiphoton absorption) and subsequent change of the electron momentum as described by the Kapitza-Dirac effect[19–22]. This effect describes the diffraction of an electron from a standing-wave light pattern formed by two counter propagating light beams. Subsequently, the electron gets diffracted in transverse momentum orders separated by $2k_{\rm ph}$, where $k_{\rm ph}$ is the carried momentum of the counter-propagating light beams. Beyond this particle point of view, it is also possible to explain this elastic interaction by the scattering from the ponderomotive potential formed by the standing wave[23]. The Kapitza-Dirac effect has also been investigated using two different laser beam wavelengths. If an electron interacts with these two light fields, energy conservation can occur through a multiphoton process in which $N$ photons of $\omega_1$ are absorbed and $M$ photons of $\omega_2$ are emitted, satisfying the condition $N\omega_1 = M\omega_2$, so that the interaction remains elastic. Notably, when this condition holds with $\omega_1 = 2\omega_2$, the absorption of two $\omega_1$ photon and the emission of one $\omega_2$ photon can conserve energy while also manipulating the electron momentum by $4k_{\rm ph}$[24]. The control of electron momentum states within the Kapitza-Dirac effect has also been generalized for quantum coherent control, allowing the preparation of arbitrary transverse momentum states through quantum path interferences[25].

The inelastic analogue to the Kapitza-Dirac effect is inelastic stimulated Compton scattering. Using two light beams with different colors, specified by the frequencies $\omega_1$ and $\omega_2$ and wave vectors $\vec{k}_1$ and $\vec{k}_2$, and considering the non-recoil approximation, the criterion for having inelastic scattering is given as $\omega_1 - \omega_2 = \vec{v}_{\rm el} \cdot (\vec{k}_1 - \vec{k}_2)$[26,27]. The quantized energy gain is accompanied with the absorption of one $\omega_1$ photon and the emission of one $\omega_2$ photon, while the energy loss corresponds to the opposite situation[7,28,29]. This inelastic manipulation of the electron wavepacket has also been demonstrated experimentally in free space by utilizing two inclined laser beams with different wavelengths, causing a dispersive propagation of the electron wavepacket in the ponderomotive potential landscape generated by the two optical waves. This inelastic manipulation leads to an acceleration of the electron[30], and the formation of distinct energy sidebands at multiples of $\hbar(\omega_1 - \omega_2)$[29]. This can be furthermore applied for the generation of attosecond electron bunches in free space[31]. In the Kapitza-Dirac effect, the two beams share the same frequency and are non-inclined, resulting in identical momenta. This characteristic prevents any inelastic electron light interaction. Recently it was demonstrated by us that it is not necessary to have two laser beams to achieve inelastic electron scattering from light with the subsequent formation of district energy sidebands[32]. This interaction is enabled using a pulsed structured light beam which supports the inelastic scattering from the pulse envelope and the self-interference of the electron within the ponderomotive landscape of the structured light field.

In this work we demonstrate numerically the simultaneous elastic and inelastic scattering of an electron wavepacket at a structured-light grating. Therefore, we systematically investigate the Kapitza-Dirac effect under consideration of spatial and temporal structure to the counterpropagating light fields. Thereby we outline the influence of spatial and temporal structure separately and show that the combination of spatially structured pulsed counterpropagating laser beams can introduce a comb like electron energy modulation in the diffraction orders attributed to the Kapitza-Dirac effect. Furthermore, we outline a strategy to utilize this interaction to achieve a controlled transverse and longitudinal phase modulation of the electron wavepacket.

**Results**

We start by considering an electron wavepacket $\psi(\vec{r}, t)$ with initial velocity $v_{el}$ traveling along the *x*-axis, and encountering two counterpropagating light waves of frequency $\omega$, wave vector $\vec{k}$, and pulse duration $\Delta\tau$ (full width at half maximum (FWHM)), along the *y*-axis (Fig. 1a). The amplitudes of these time-dependent electromagnetic fields are represented by the power normalized Hermite-Gaussian (HG) modes which are solutions to the paraxial Helmholtz equation (see Methods and supplementary note 1). Especially we will focus on the HG modes of order HG$_{00}$ and HG$_{10}$. Applying this, one obtains the following expression for the time dependent vectorpotential in two dimensions[33],

$$\vec{A}_n(\vec{r},t) = \vec{y} \frac{A_0}{N_{P,n}} \frac{w_0}{W(y)} H_n\left(\frac{\sqrt{2}x}{W(y)}\right) \exp\left(-\frac{x^2}{W(y)^2}\right)$$
$$\times \exp\left(i\left[|\vec{k}|y - (1+n)\tan^{-1}\frac{y}{y_0} + \frac{|\vec{k}|x^2}{2R(y)}\right]\right) \exp\left(-i\omega t - \frac{(t-t_0)^2}{4\Delta\tau^2}\right) \quad (1)$$

where $A_0$ is the amplitude of the vectorpotential, $N_{P,n}$ is the order dependent power normalization constant and $\omega = 2\pi c/\lambda$ is the angular frequency of the light wave. Here, $W(y) = w_0\sqrt{1+(y/y_r)^2}$ with $w_0$ and $y_r = \pi w_0^2/\lambda$ denoting the beam waist and Rayleigh range, respectively. The radius of curvature is given by $R(y) = y[1 + (y/y_r)^2]$ and $H_n$ are the Hermite polynomials of order n. The power normalization leads to different peak amplitudes for the different HG$_{00}$ and HG$_{10}$ modes. $t_0$, the temporal center, synchronizes the arrival time of the optical pulse with the electron wavepacket in the time domain. In this work we set the amplitude $A_{0,0}$ of the HG$_{00}$ mode as reference and determine the amplitude $A_{1,0}$ of the HG$_{10}$ beams with $A_{1,0} = \sqrt{2/e}\, A_{0,0}$ (see methods for derivation).

We further confirm whether the two counterpropagating waves given by equation (1) can fulfill the condition of inelastic stimulated Compton scattering. Therefore, it is useful to expand the higher-order Gaussian beams as a superposition of plane waves[34]:

$$\vec{A}_n(\vec{r}) = \vec{e}_y \int A_n(k_\perp) \, e^{i\vec{k}\cdot\vec{r}} dk_\perp \quad , \quad (2)$$

where the wave vector of the light is given by $\vec{k} = k_\perp \vec{e}_x + k_\parallel \vec{e}_y$ with $k = \sqrt{k_\perp^2 + k_\parallel^2}$. The vector potential is therefore recast as a superposition of plane waves propagating at the angles of $\theta_k = \tan^{-1}(k_\perp/k_\parallel)$ with respect to the beam y-axis. The amplitude $A_n(k_\perp)$ represents the transverse momentum distribution within the beam and can be interpreted as the Fourier transform of the amplitude distribution $A_n(x, y)$ given by

$$A_n(k_\perp) = \frac{1}{\sqrt{2\pi}} \int A_n(x, y=0) e^{-ik_\perp x} dx \quad . \quad (3)$$

Here it becomes clear that these higher-order Gaussian beams are a superposition of plane waves with different transverse momenta and therefore, different opening angles with respect to the beam axis. Fig. 1b displays the transverse momentum distribution for the higher-order Gaussian beams of the order 00 and 10. Consequently two spatially structured and counter propagating light waves with the same central frequency ω, carry a non-zero photon momentum distributions in the propagation direction of the electron wavepacket. This allows to recast the condition of inelastic stimulated Compton scattering as following:

$$\omega_1 - \omega_2 = v_{el}(k_{\perp,1} - k_{\perp,2}) \, . \quad (4)$$

This condition can be fulfilled in the case of short pulse durations with a broad spectrum centered around ω, and electron velocities in the range of 1 to 30 keV (see Fig. 1b upper panel and supplementary note 2). Consequently, the electron can gain, and loose energy in orders of $\hbar v_{el}(k_{\perp,1} - k_{\perp,2})$ (Fig. 1c). In addition to the inelastic stimulated Compton scattering, also the condition of elastic stimulated Compton scattering is fulfilled by the wavevectors in the direction of light beam propagation. These are given by $k_{\parallel,1}$ and $k_{\parallel,2}$. This fact results in the electron wavepacket being diffracted into the orders determined by $2k_{\parallel,2}$ (when $k_1 = k_2$) or $4k_{\parallel,2}$ (when $k_1 = 2k_2$) (Fig. 1d). The combination of these effects leads to a final electron momentum distribution demonstrating a manipulation of both transverse and longitudinal momenta (Fig. 1e).

To further discuss the physics of this interaction we choose two approaches. Analytically the dynamics of such free-space electron wavepackets in light gratings are commonly described in terms of the Volkov representation of the electron wavefunction propagating through a vector potential $A(\vec{r}, t)$. This method has proven especially useful in describing the Kapitza-Dirac effect[25]. However, it is important to note that the Volkov wave function alone does not fully describe all conceivable electron-light interactions, particularly in the case of slow-electrons where the slowly varying approximation may be inadequate. Additionally, the Volkov approximation is limited by the dipole approximation (neglecting the spatial distribution of light) and may prove inaccurate when accounting for the spatial structure of electromagnetic fields. Therefore, we will additionally discuss this problem with the help of a previously developed numerical toolbox that solves the time-dependent Schrödinger equation within the minimal-coupling Hamiltonian formalism combined with a numerical method to solve Maxwell's equations to obtain the electromagnetic field (see methods). In these numerical calculations the electron wavepacket is described by a Gaussian envelope with longitudinal ($W_L$) and transverse ($W_T$) broadenings at FWHM. The considered structure for the electromagnetic fields with which the electron wavepacket interacts are two counterpropagating HG pulsed beams with the center wavelength λ and beam waist $w_0$. These electromagnetic fields are introduced within the finite-difference time-domain-based Maxwell solver by a Hermite-Gaussian field profile. In the following sections, we will analyze the underlying interaction mechanisms and progressively illustrate how structured and pulsed light manipulate the electron momentum distribution within the context of stimulated Compton scattering.

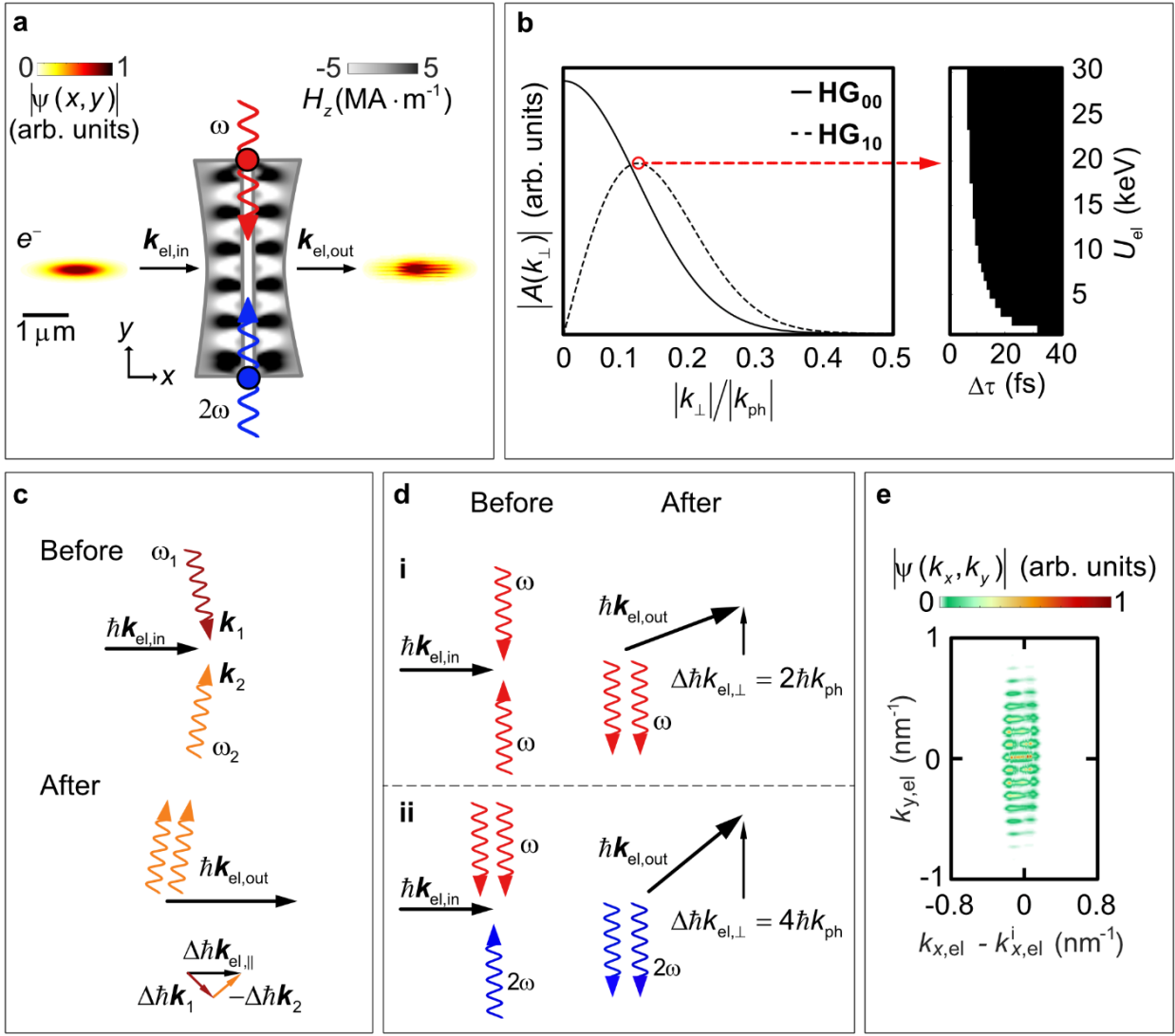

**Fig 1. Inelastic and elastic electron scattering from counter propagating spatially structured pulsed optical beams.** **a**, Illustration of an electron wavepacket traversing two counter propagating pulsed structured light beams of the frequencies $\omega$ and $2\omega$. **b**, Transverse momentum distribution of higher-order gaussian modes (left panel) and fulfillment of the inelastic stimulated Compton scattering condition (white=true; black=false) for a Hermite-Gaussian (HG) 10 beam of the wavelength $\lambda = 500$ nm versus pulse duration and electron velocity ($k_{\mathrm{ph}} = \omega/c$). **c**, Illustration of inelastic stimulated Compton scattering (energy gain). **d**, Illustration of elastic stimulated Compton scattering. (i) from two photons of the same frequency $\omega$ (ii) from two photons of frequency $\omega$ and one photon of frequency $2\omega$. **e**, Final electron momentum distribution of the electron wavepacket after the interaction. The considered electron has an initial kinetic energy of 1 keV and transverse and longitudinal full width at half maximum (FWHM) of $W_L = 250$ nm and $W_T = 60$ nm respectively. The

counter propagating light fields have a HG$_{10}$ amplitude distribution and wavelengths of $\lambda = 300$ nm and $\lambda = 150$ nm as well as pulse FWHM of $\Delta\tau = 10$ fs. The field amplitude was set to 15 GVm$^{-1}$.

**Kapitza-Dirac effect for a spatially structured standing-wave light pattern**

The Kapitza-Dirac effect can be analyzed in either the Bragg or diffraction regime. The dominance of one regime over the other depends on the spatial extent of the light field with which the electrons interact and on the duration of the electron-light interaction[22]. This spatial extent of the optical beam however is not a determining factor for shaping the electron beam with velocities commonly used in electron microscopes (see supplementary note 3). However, despite its spatial extent, the fine spatial structure of the light may influence elastic interaction processes, owing to the higher momentum orders introduced by such spatially-structured light (Fig. 1b). Therefore, it is interesting to further discuss a possible influence of the spatial structure of light on the diffraction orders of the Kapitza-Dirac effect. For this we first consider an analytical approach to this problem by comparatively calculating the Volkov wave function for the continuous-wave (CW) standing-wave light patterns with HG$_{10}$ and HG$_{00}$ orders (see methods). This discussion results into two identical diffraction probability expressions, suggesting that the Kapitza-Dirac effect is independent of the spatial mode structure of the considered light field. To support this claim, we obtain numerical calculations for the two different spatial modes. The calculations consider a Gaussian electron wavepacket with the center kinetic energy of 1 keV and $W_\mathrm{L} = 250$ nm, $W_\mathrm{T} = 60$ nm interacting with a standing-wave light field of the wavelength $\lambda = 300$ nm and beam waist $w_0 = 2\lambda$. The amplitude of HG$_{00}$ and HG$_{10}$ modes, are 5 GVm$^{-1}$ or 4.3 GVm$^{-1}$, respectively. The resulting transverse momentum distributions of the electron pulse after the interaction with the HG$_{10}$ and HG$_{00}$ standing-wave light field demonstrate almost no difference (Fig. 2a). The observed absolute probability amplitude exhibits only slight differences. The high population probability towards higher transverse momentum orders as observed in Fig. 2a is a well-known phenomenon for electron-light interactions in strong electromagnetic fields[35,36]. The overall probability distribution demonstrates the shape characteristic of a quantum walk due to quantum path interferences which are also expected at these high intensities[25]. Thus, these calculations confirm the analytical conclusions that the Kapitza-Dirac effect does not differ for higher-order spatially structured standing light waves. However, arriving at a definitive conclusion regarding whether the introduction of spatial structure results in distinct solutions compared to the conventional planewave-based solution remains challenging based on the analytical and numerical results obtained so far. Therefore, we add further calculations showing the influence of the beam waist $w_0$ on the Kapitza-Dirac effect (as displayed in Fig. 2b). In one calculation the beam waist is selected to one extremely narrow case of $w_{0,1} = 1.5\lambda$ which is associated with a large opening angles and thus differs strongly from a plane wave description. The other calculation is based on a wider beam waist of $w_{0,2} = 4\lambda$, consequetly having a smaller opening angle distribution. One observes no difference between the results. These results overall leads to the conclusion that the spatial structure of light has no influence on the Kapitza-Dirac effect in CW standing-wave light patterns within the diffraction regime.

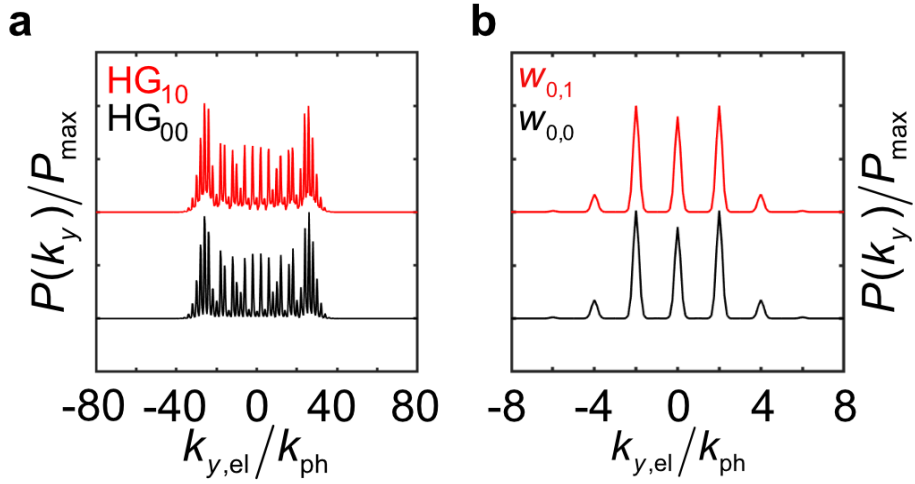

**Fig. 2. Kapitza-Dirac effect with spatially-structured light. a**, Transverse momentum distribution of the electron pulse after the interaction with two different Hermite-Gaussian (HG) (HG$_{10}$ (red) and HG$_{00}$ (black)) standing-wave light fields ($k_{\mathrm{ph}} = k_2$). The CW electric field amplitudes of the spatial modes are 5 GVm$^{-1}$ (HG$_{00}$) and 4.3 GVm$^{-1}$ (HG$_{10}$), respectively. All light fields have a wavelength of $\lambda = 300$ nm. **b**, Transverse momentum distribution of the electron pulse after the interaction with a HG$_{00}$ standing-wave light field with the beam waists of $w_{0,1} = 1.5\lambda$ and $w_{0,2} = 4\lambda$. The amplitude of the CW electric field and the wavelength are 50 GVm$^{-1}$ and $\lambda = 150$ nm, respectively. The electron wavepacket has an initial kinetic energy of 1 keV and longitudinal and transverse full width at half maximum (FWHM) of $W_{\mathrm{L}} = 250$ nm and $W_{\mathrm{T}} = 60$ nm, respectively in **a** and **b**. In all plots, the momentum distributions ($P(k_y)$) are normalized to the maximum observed probability amplitude ($P_{\mathrm{max}}$) and placed with an offset relative to each other.

**Kapitza-Dirac effect for a pulsed standing-wave light pattern**

Next, we discuss the influence of pulsed light fields on the Kapitza-Dirac effect. In this case the analytical discussion is based on the two counter-propagating plane waves which are additionally in the form of a Gaussian-shaped pulse with the pulse duration $\Delta\tau$. The probability distribution obtained from the Volkov representation (see methods equation (24)) is displayed in Fig. 3a. These distributions demonstrate that the populated diffraction orders depend on the pulse duration such that for longer pulse durations the probability to populate higher diffraction orders is larger. Additionally, with increasing the field strength it is further possible to enhance this trend toward the population of even higher diffraction orders. The observation that shorter pulse durations lead to a decrease in the number of populated transverse diffraction orders was made during the first observation of the Kapitza-Dirac effect with sodium atoms[36]. To explain this effect, it was argued that the number of populated diffraction orders solely depends on the field intensity and the interaction time of the matter wave with the standing light wave. For a pulsed beam, the interaction time changes according to the pulse duration. Thus, for very short pulses, the electron-light interaction strength becomes weaker.

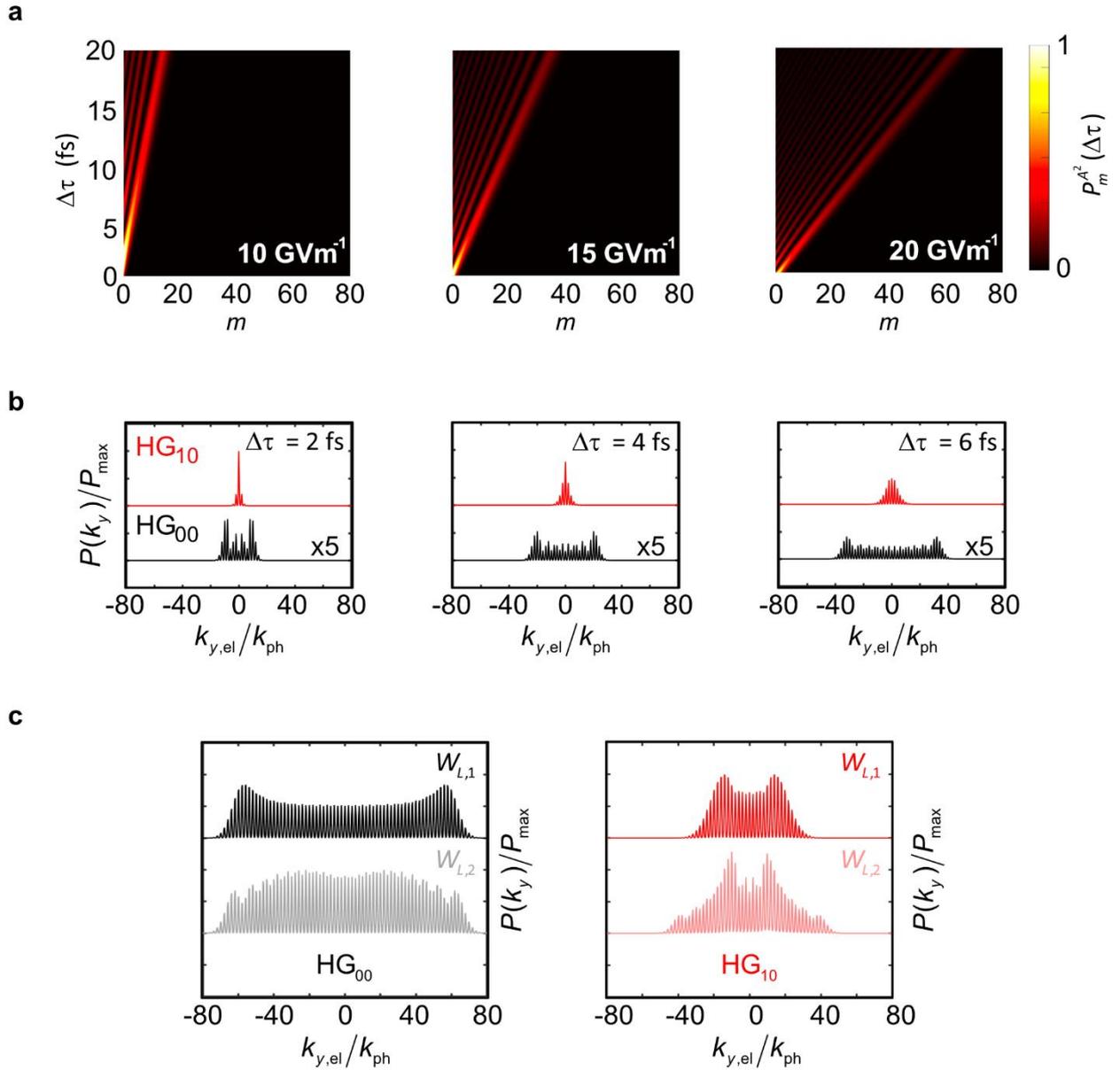

**Fig. 3. Influence of pulsed beams on the Kapitza-Dirac effect. a**, Analytical probability distribution for an electron populating the $m^{th}$ diffraction order after interacting with a standing wave pattern with a pulse duration of $\Delta\tau$ and wavelength of 300 nm for the electric field amplitudes of 10 GVm$^{-1}$, 15 GVm$^{-1}$ and 20 GVm$^{-1}$. **b**, Numerically obtained transverse momentum spectrum after the interaction with two counter propagating Hermite-Gaussian (HG) beams ($k_{\text{ph}} = k_2$). The subsequently HG$_{10}$ (red) and HG$_{00}$ (black) structured standing wave patterns have the variation of the pulse duration from $\Delta\tau = 2$ fs (left), $\Delta\tau = 4$ fs (middle) to $\Delta\tau = 6$ fs (right). The transverse momentum spectra after the interaction with a HG$_{00}$ beam is magnified by a factor of 5. The electron wavepacket has a longitudinal full width at half maximum (FWHM) of $W_{\text{L}} = 50$ nm. **c**, Final transverse momentum spectra for an electron wavepacket with the spatial longitudinal FWHM of $W_{\text{L},1} = 50$ nm or $W_{\text{L},2} = 250$ nm. This is depicted for transverse momentum spectra obtained after the interaction with a HG$_{00}$ (left panel) and a HG$_{10}$ (right panel) beam. A pulse duration of 10 fs is chosen. In b and c the considered light fields have the electric-field-amplitude of 10 GVm$^{-1}$ (HG$_{00}$) and 8.6

GVm$^{-1}$ (HG$_{10}$), respectively. The wavelength in all cases is $\lambda = 300$ nm. The electron wavepacket has an initial kinetic energy of 1 keV in all calculations. In panels **b** and **c**, the momentum distributions are normalized to the maximum observed probability amplitude and placed with an offset relative to each other.

Using the numerical Maxwell-Schrödinger framework, we now systematically explore the above-mentioned effects beyond the dipole approximation. For these calculations we choose an electron wavepacket having a longitudinal FWHM of $W_L = 50$ nm and center kinetic energy of 1 keV interacting with two counterpropagating pulsed HG$_{00}$ beams with the central wavelength of $\lambda = 300$ nm, and varied pulse duration of $\Delta\tau = 2$ fs, $\Delta\tau = 4$ fs and $\Delta\tau = 6$ fs. The beam waist is set to $w_0 = 2\lambda$ and the considered light fields have the electric-field amplitude of 10 GVm$^{-1}$ (HG$_{00}$) and 8.6 GVm$^{-1}$ (HG$_{10}$). The calculations for varied pulse durations (Fig. 3b) reveal that for a shorter pulse duration the number of populated transverse diffraction orders (separation by $2k_{\parallel,2}$) decreases as predicted by our analytical calculations. This effect is observed for both the HG$_{00}$ and HG$_{10}$ standing-wave light field. The final transverse momentum spectrum of the electron interacting with the HG$_{00}$ standing-wave light field thereby consistently shows a higher number of populated momentum orders. When modifying the spatial structure of the light by introducing a HG$_{10}$ amplitude, the interacting electron experiences a reduced amplitude throughout the same pulse duration (see supplementary note 4). Consequently, this further weakens the interaction strength. Provided that the optical pulse duration is long enough, the electron wavepacket can interact with the complete field profile of the HG$_{10}$ and HG$_{00}$ standing-wave light fields. Given this condition, the transverse momentum distribution should show no differences between those two interaction geometries. This situation is similar to the results obtained for the electron interacting with a CW standing-wave light field (see Fig. 2). It is worth noting that the probability distribution of the transversely populated momentum states for the interaction with the HG$_{00}$ show variations in the probability amplitudes for different diffraction orders. This can be explained through quantum-path interferences, that become specially relevant at higher intensities[25].

Next, we consider the influence of the longitudinal broadening of the electron wavepacket on the interaction strength. For this a large spatial extent of $W_L = 250$ nm is considered. It was already observed that a large spatial extent of the electron wavepacket strongly influences the final energy-gain spectrum[32]. The results for both HG$_{00}$ and HG$_{10}$ spatial modes are then compared to the calculations with an electron wavepacket of $W_L = 50$ nm. For these calculations, furthermore the optical pulse duration was fixed to 10 fs. The resulting transverse momentum spectra (Fig. 3c) show a varied probability distribution in the final momentum distribution for a larger electron wavepacket. In the case of an interaction with a HG$_{00}$ standing-wave light pattern, the overall number of populated diffraction orders remains unchanged. For a HG$_{10}$ standing-wave light pattern, a different result is observed. In this case, the probability distributions exhibit the population of higher transverse momentum orders for the electron of larger spatial extent. This observation indicates that the large longitudinal extent $W_L$, of the electron wavepacket influences the final electron transverse momentum distribution for a pulsed standing-wave light field. Specifically, in the case of the HG$_{10}$ geometry employed here, one observes an enhanced electron-light interaction causing the population of more diffraction orders. This can be attributed to the larger interaction area of a larger electron wavepacket with the beam profile of the HG$_{10}$ mode. The same applies to the longer electron wavepacket interacting with the HG$_{00}$ beam. Yet, in the case of a 10 fs light pulse, the electron has already encountered almost the complete beam profile due to the spatially confined geometry of the HG$_{00}$ mode. Consequently, the number of populated diffraction orders is not affected. However, the overall shape of the spectra is significantly altered.

Finally, we observe the suppression of additional features in the final electron momentum spectrum for $W_L = 50$ nm and the interaction with both HG$_{00}$ and HG$_{10}$ standing-wave light fields. We previously attributed these features to be introduce by quantum path Interferences. The size of the electron wavepacket strongly affects the ability to interfere with itself and therefore such interference effects usually are known to be suppressed for a shorter electron wavepacket[25,37] (see also supplementary note 5).

**Inelastic and elastic electron scattering from counterpropagating spatially-structured pulsed light waves**

Up to this point, we have explored the impact of the spatial structure and finite pulse duration on the transverse momentum distribution in the Kapitza-Dirac effect. In particular, we will focus on the observed final energy modulations. Thus, the simulation parameters for both the HG$_{00}$ and HG$_{10}$ systems are chosen accurately to ensure that the interaction falls within the regime of inelastic stimulated Compton scattering. Therefore, we investigate the interaction of an electron wavepacket with $W_L = 250$ nm, $W_T = 60$ nm and the center kinetic energy of 1 keV with two counterpropagating pulsed optical beams of the same HG spatial mode (HG$_{10}$ and HG$_{00}$) with central wavelength of $\lambda = 300$ nm, pulse duration of $\Delta\tau = 10$ fs, beam waist of $w_0 = 2\lambda$ and electric field amplitudes of 5 GVm$^{-1}$ (HG$_{00}$) and 4.3 GVm$^{-1}$ (HG$_{10}$).

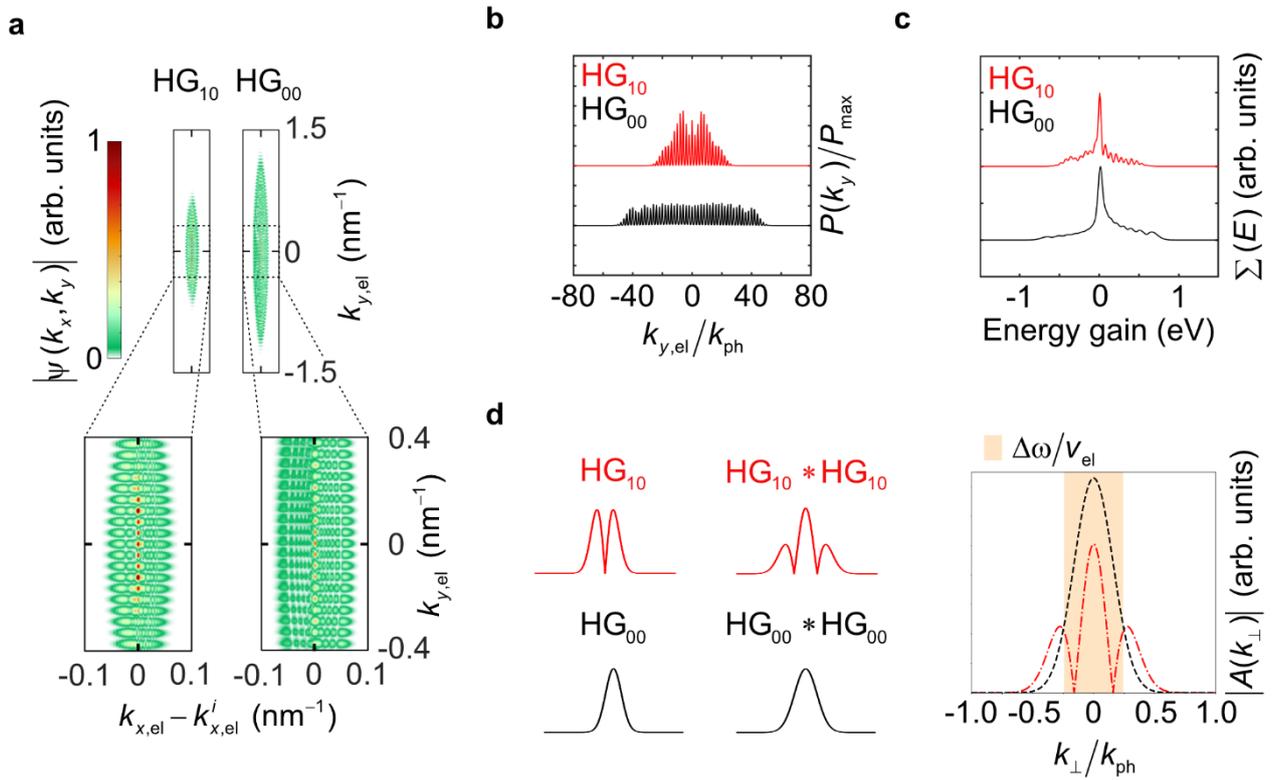

**Fig. 4. Elastic and inelastic electron scattering.** Elastic and inelastic scattering of an electron wavepacket by pulsed and spatially structured (Hermite-Gaussian (HG)) standing-wave light patterns. The optical pulses have a temporal full width at half maximum (FWHM) of 10 fs and the standing-wave light fields have either an HG$_{10}$ or HG$_{00}$ spatial structure. **a**, Depiction of the transverse and longitudinal momentum distribution for all populated final momentum states and close-up of the final transverse and longitudinal momentum

population. **b**, Final transverse momentum distribution in orders of the photon momentum after the interaction with the HG$_{10}$ (top, red) and HG$_{00}$ (bottom, black) standing-wave light pattern ($k_{\text{ph}} = k_2$). **c**, Final electron energy gain spectrum after the interaction with the HG$_{10}$ (top, red) and HG$_{00}$ (bottom, black) standing light waves. The considered light fields have the electric-field-amplitude, wavelength, and temporal broadening of 5 GVm$^{-1}$ (HG$_{00}$) and 4.3 GVm$^{-1}$ (HG$_{10}$), 300 nm and 10 fs, respectively. The electron wavepacket has an initial kinetic energy of 1 keV, and initial longitudinal and transverse FWHM of $W_{\text{L}} = 250$ nm and $W_{\text{T}} = 60$ nm, respectively. In **b** and **c** the momentum distributions and electron energy gain spectra are normalized to the maximum observed probability amplitude. **d**, Profile of effective transverse momentum distributions, obtained by convolving superimposed Hermite-Gaussian beams (left panel). The orange color denotes fulfillment of inelastic stimulated Compton scattering under the considered interaction parameters (right panel).

The first aspect demonstrated by these simulations is the population of final longitudinal and transverse momentum states, resulting in an eye-like-shaped electron distribution in the momentum representation (Fig. 4a). The population of the transverse momentum orders is determined by the mechanisms outlined in the previous section and depends strongly on the pulse duration and the spatial mode of the light beams and leads to the populated momentum orders separated by $2k_{\parallel,2}$ (Fig. 4b). The inelastic scattering from the pulsed spatially-structured standing-wave light patterns results for this geometry in two similar electron energy gain spectra for both spatial modes. The electron energy-gain spectra in Fig. 4c demonstrates a dominant zero-loss peak and the formation of weakly visible sidebands towards higher gain and loss channels in the final electron energy spectrum. It is noticeable that the electron energy spectrum obtained after the interaction with the HG$_{10}$ spatial modes demonstrate a less pronounced longitudinal extent compared to the interaction with the HG$_{00}$ spatial modes. This is again explainable trough the reduced amplitude that the electron experiences throughout the interaction with HG$_{10}$ spatial modes at finite pulse durations. Additionally, it is observable that the electron energy-gain spectrum has a larger number of distinct sidebands.

The process of stimulated Compton scattering initiated by two optical beams is caused by a simultaneous stimulated emission and absorption of photons by the electron. This requires the vectorial addition of the momentum components of the respective beams as described in Fig. 1b. Similarly, the ponderomotive potential formed by the two beams is recast as

$$|\vec{A}|^2 = |A_z|^2 = \frac{E_{01}^2}{\omega_1^2}\cos^2(\vec{k}_1 \cdot \vec{r} - \omega_1 t) + \frac{E_{02}^2}{\omega_2^2}\cos^2(\vec{k}_2 \cdot \vec{r} - \omega_2 t), \quad (5)$$
$$+2\frac{E_{01}}{\omega_1}\frac{E_{02}}{\omega_2}\cos(\vec{k}_1 \cdot \vec{r} - \omega_1 t)\cos(\vec{k}_2 \cdot \vec{r} - \omega_2 t)$$

where the last term includes the multiplication of the spatial distributions of the interacting waves. Hence, in the momentum domain, this translates to the convolution of the momentum distributions of the optical beams, as shown in Fig. 4d. The momentum distribution of the resulting ponderomotive potential therefor constitutes the required momentum for initiating the stimulated Compton scattering by interacting with the electron beam, that further results in the longitudinal energy sidebands as observed in the simulations.

A detailed analysis shows that these sidebands are fluctuating in their respective distance and that their average distance is given by $\Delta E = 73 \pm 3$ meV while their median is at $\Delta E = 63$ meV. The fluctuations was obtained as the standard error $\sigma_{\bar{k}} = \sigma/\sqrt{N}$, where σ is the standard deviation of the sideband distribution, and N the number of considered sidebands. This approach of determining the deviation will stay consistent for the coming discussions. This average change in energy corresponds to the stimulated interaction with two photons with the wavelength of $\lambda_1 = 302.5$ nm and $\lambda_2 = 297$ nm, which their difference is within the spectral range of the introduced Gaussian laser pulse. As mentioned above, their respective weight in the interaction can be understood by the convolution of the two transverse momentum distributions (see methods and Fig 4d). This results in a modulated transverse momentum distribution, demonstrating a pronounced narrow maximum around zero, and additionally two less populated symmetric maxima. Two photons that carry a transverse momentum of $\pm k_\perp$, which is in the order of magnitude given by the FWHM of the central maximum around the zero point of the distribution, can fulfill equation (4) and would lead to an energy gain of $\Delta E = 2\hbar v_{el} k_\perp = 61$ meV. This is close to the mean of the observed sideband distribution, whose fluctuation is explainable by the continuous distribution of momentum components that as well satisfy the criterium of stimulated Compton scattering. With the bandwidth of the laser pulse allowing a maximum energy gain or loss of $\Delta E = \hbar(\omega_1 - \omega_2) = 130$ meV, not the complete transverse momentum distribution satisfies equation (4). Nevertheless, the high transverse momentum components of the distribution that allows an inelastic interaction according to equation (4), can lead to stimulated interactions that result in a higher energy gain or loss process, thus shifting the average and median sideband distance.

The electron energy spectrum obtained after the interaction with the HG$_{00}$ spatial modes also demonstrates the formation of energy sidebands. These sidebands have as well, a varying energy separation. For the lower diffraction orders (*m*<10) the average of this separation is given by $\Delta E = 108 \pm 4$ meV. The median separation is $\Delta E = 102$ meV. To explain this observation, it is again useful to determine the convoluted transverse momentum distribution of the HG$_{00}$ standing-wave light pattern, which results in a more broadened (compared to the distribution of a single HG$_{00}$ beam, see supplementary note 2) Gaussian-shaped transverse momentum distribution (see methods and Fig. 4d). Both statistically obtained sideband separations are close to the energy gain supported form the transverse momentum states from the region around the FWHM of this Gaussian-shaped transverse momentum distribution. The stimulated interaction with two photons carrying a transverse momentum of $\pm k_\perp$ from that region would lead to an energy gain of $\Delta E = 2\hbar v_{el} k_\perp = 108$ meV. The deviation of median and average are again explainable by the continues transverse momentum distribution. However, in contrast to the previous discussion we observe for higher diffraction orders (*m*>10) an increase in the sideband separation reaching energy gain and losses higher than allowed ($\Delta E > \hbar(\omega_1 - \omega_2) = 130$ meV) by equation (4). In this region of the final electron momentum spectrum, we observe an average sideband distribution of $\Delta E = 141 \pm 5$ meV and a median distribution of $\Delta E = 136$ meV. A possible explanation is that the elevated peak intensities of an HG$_{00}$ standing-wave light pattern enable stimulated interactions involving multiple photons *N* carrying the same energy. This would lead to the formation of energy gain and loss peaks separated by $\Delta E = N\hbar(\omega_1 - \omega_2)$. Such higher order nonlinear interactions have been introduced for both elastic and inelastic stimulated Compton scattering[24,38].

So far, we have not discussed the possible origin of the observed dominant zero-loss peak. This observation can be discussed in relation to the self-interference of the electron wavepacket during the scattering from a single pulsed structured light wave. For this interaction, the formation of the amplitude of the final electron energy gain spectrum was explained by electron wavepacket self-interference[32]. The interaction dynamics for the Kapitza-Dirac effect for both the $HG_{10}$ and $HG_{00}$ waves demonstrate no longitudinal oscillation pattern during the interaction as observed for the self-interference effect (see supplementary note 6). This suggests that the ponderomotive force acting along the longitudinal axis of the standing-wave light pattern with $HG_{10}$ and $HG_{00}$ shapes differs from the force exerted by the potential of a traveling beam (see supplementary 7). This prevents any electron energy modulation due to the self-interference. For larger field intensities the interaction might enter the strong-coupling regime[35,39,40] for the longitudinal momentum modulation which results in higher energy orders being predominantly populated. A variation of the respective pulse duration of the two beams nicely demonstrates that the temporal overlap of both beams should be matched to achieve the highest energy gain as well as a simultaneous strong elastic interaction (see supplementary note 8). Furthermore, it is possible to discuss the interaction in terms of varying pulse duration and field intensity. For this we consider two interactions with two counter propagating $HG_{10}$ beams (see supplementary note 9). In the first scenario both beams have a pulse duration and field intensity of $\Delta\tau = 10$ fs and 15 GVm$^{-1}$, respectively. The second scenario considers a pulse duration and field intensity of $\Delta\tau = 20$ fs and 10 GVm$^{-1}$ respectively. These calculations lead to the conclusion that an increase in the field intensity overall strengthens the interaction. The choice of longer pulse durations, which essentially translates to an increased interaction time, does not affect the strength of the inelastic scattering. However, the number of populated transverse momentum states increases for longer pulse durations and weaker field intensities. Moreover, our results illustrate that elastic stimulated Compton scattering, exhibits a higher interaction strength than inelastic stimulated Compton scattering, reaching the strong interaction regime at lower field intensities. Furthermore, it should be mentioned that the electron velocity influences the interaction strength as well. By increasing the electron velocity (still considering the fulfillment of equation (4)) the overall interaction strength reduces resulting into the population of fewer transverse and longitudinal momentum orders (see supplementary Note 10).

**Control of inelastic and elastic electron scattering by pulsed spatially structured light waves**

Altogether these results lead to the question of how one might be able to control the final electron momentum distribution by combining different spatial modes at varying wavelengths. To systematically address this question, we first alter the previous systems by mixing both the $HG_{10}$ and $HG_{00}$ mode during the interaction (see Fig. 5 a, d, g, j). Each beam has a wavelength of $\lambda = 300$ nm and a pulse duration of $\Delta\tau = 10$ fs. The electric field amplitudes are 5 GVm$^{-1}$ ($HG_{00}$) and 4.3 GVm$^{-1}$ ($HG_{10}$).

This results into a final electron momentum spectrum of a rhombus-like shape (Fig. 5d). The transverse momentum population is determined by a momentum order separation of $2k_{\parallel,2}$, demonstrating the highest probability for the 0th order diffraction peak (Fig. 5g). This transverse momentum distribution indicates that the interaction is not within the strong-coupling regime. This at the same time comes with the absence of a probability modulation due to the quantum-path interferences. The final electron energy spectrum observed at the different diffraction orders however shows a drastic difference compared to the calculations with superimposed identical spatial mode (Fig. 5j). The dominant zero loss peak is not observable anymore and the electron energy gain spectrum demonstrates the population of high probability states towards higher energies. The spectral shape is strongly asymmetric, in contrast with generally symmetric PINEM spectra within the weak-interaction regime. The interaction dynamics (see supplementary note 11) demonstrate the appearance of longitudinal oscillatory dynamics within the electron momentum distribution. These oscillatory dynamics are caused by the ponderomotive landscape formed by the two different counterpropagating beams. This phenomenon therefore influences the final modulation of the probability amplitude. The separation of the diffraction orders of the electron energy spectrum demonstrates a large variety with an average at $\Delta E = 86 \pm 9$ meV. To understand this spacing, it is useful to again discuss the superposition of the transverse momentum distribution of the $HG_{10}$ and $HG_{00}$ spatial modes by performing a convolution of these two distributions (see methods). The resulting distribution has a similar but more broadened shape compared to the one obtained for a single $HG_{10}$ beam (see supplementary note 2 and 7). This broadening can explain the increased sideband peak-position fluctuations. Thus, the distribution also demonstrates two symmetric maxima at a distinct $\pm k_\perp^{max}$. In this case the stimulated interaction with two photons, carrying a transverse momentum of $k_\perp^{max}$ and $-k_\perp^{max}$ respectively, would lead to an energy gain of $\Delta E = 2\hbar v_{el} k_\perp^{max} = 82$ meV, being in a good agreement to the numerical result. We therefor conclude that this method is as well suited to explain the sideband separation for the interaction induced by the superposition of different spatial modes.

Next, we discuss the case of two counter-propagating beams $HG_{10}$ beams, however with the two different wavelengths of $\lambda = 300$ nm and $\lambda = 150$ nm and each with a pulse duration of $\Delta \tau = 10$ fs and electric field amplitude of 4.3 GVm$^{-1}$ (Fig. 5 b, e, h, k). The wavelengths are selected such that $\lambda_1 = 2\lambda_2$ which, within the condition of stimulated Compton scattering, leads to a momentum order separation of $4k_{\parallel,2}$ (see Fig. 5 (e,h)). The final electron momentum distribution demonstrates that the combination of two different wavelengths lead to a weaker interaction resulting in the population of fewer transverse and longitudinal momentum orders. The electron energy gain spectrum demonstrates sidebands separated by $\Delta E = 63 \pm 4$ meV (Fig. 5k). The shape of the energy gain spectrum thereby corresponds closely to the result obtained for the inelastic interaction with a single $HG_{10}$ beam with the same simulation parameters, indicating that the energy gain and loss processes are dominated by the larger wavelength. We therefore calculate the weighted transverse momentum distribution again based on the convolution of the $HG_{10}$ and $HG_{00}$ spatial mode (see methods), where we treated the wavelength $\lambda = 150$ nm as a $HG_{00}$ mode. This results into a distribution having the most likely interaction with $\Delta E = 2\hbar v_{el} k_\perp^{max} = 63$ meV, which is in a very good agreement with the numerical result.

Finally, we examine the superposition of two $HG_{00}$ modes with wavelengths of $\lambda = 300$ nm and $\lambda = 150$ nm (Fig. 5 c, f, i, l). The pulse duration and electric field amplitude are for both electromagnetic fields $\Delta \tau = 10$ fs and 5 GVm$^{-1}$, respectively. The population of transverse momentum states shows a distinct separation of $4k_{\parallel,2}$, spanning 8 diffraction orders (Fig. 5 (f, i)). This demonstrates an increased interaction strength compared to the previous calculation with the $HG_{10}$ spatial mode. This effect agrees with the previous results showing that $HG_{00}$ beams lead to stronger interactions at finite pulse durations. The electron energy gain spectra at the different diffraction orders are strongly asymmetric, showing a dominant energy gain characteristic (Fig. 5 l). This effect is also in agreement with the result obtained for the interaction with shorter electron wavepacket of $W_L = 50$ nm (see supplementary note 5). The spatial extent of the electron wavepacket is linked to the ability to coherently interact with light. Hence, in this case the electron partially interacts with the light fields similar to laser-based electron acceleration and therefore gains a finite amount of energy from the ponderomotive acceleration force[41] and only weakly interacts in the form of stimulated inelastic stimulated Compton scattering. Nevertheless, the few occurring sidebands are separated by $\Delta E = 118 \pm 18$ meV. This is a larger separation compared to the interaction involving two $HG_{00}$ beams of the same wavelength. Analyzing the transverse momentum distribution of the two beams indicates that their convolution results in a broader profile. This, in turn, leads to a statistically significant increase in the energy separation of the sidebands. However, it is essential to note that this broader distribution is accompanied by a proportionate increase in sideband distance fluctuations.

With this concluding discussion we summarize that by varying the spatial modes and wavelengths of the considered beams, it is possible to control the distance in the populated transverse momentum states and the energy separation of the observed side bands in the electron energy gain spectrum. Additionally, the ponderomotive landscape formed by the superposition of the different spatial modes can support electron wavepacket self-interference which influences the population of energy states in the final electron energy gain spectrum.

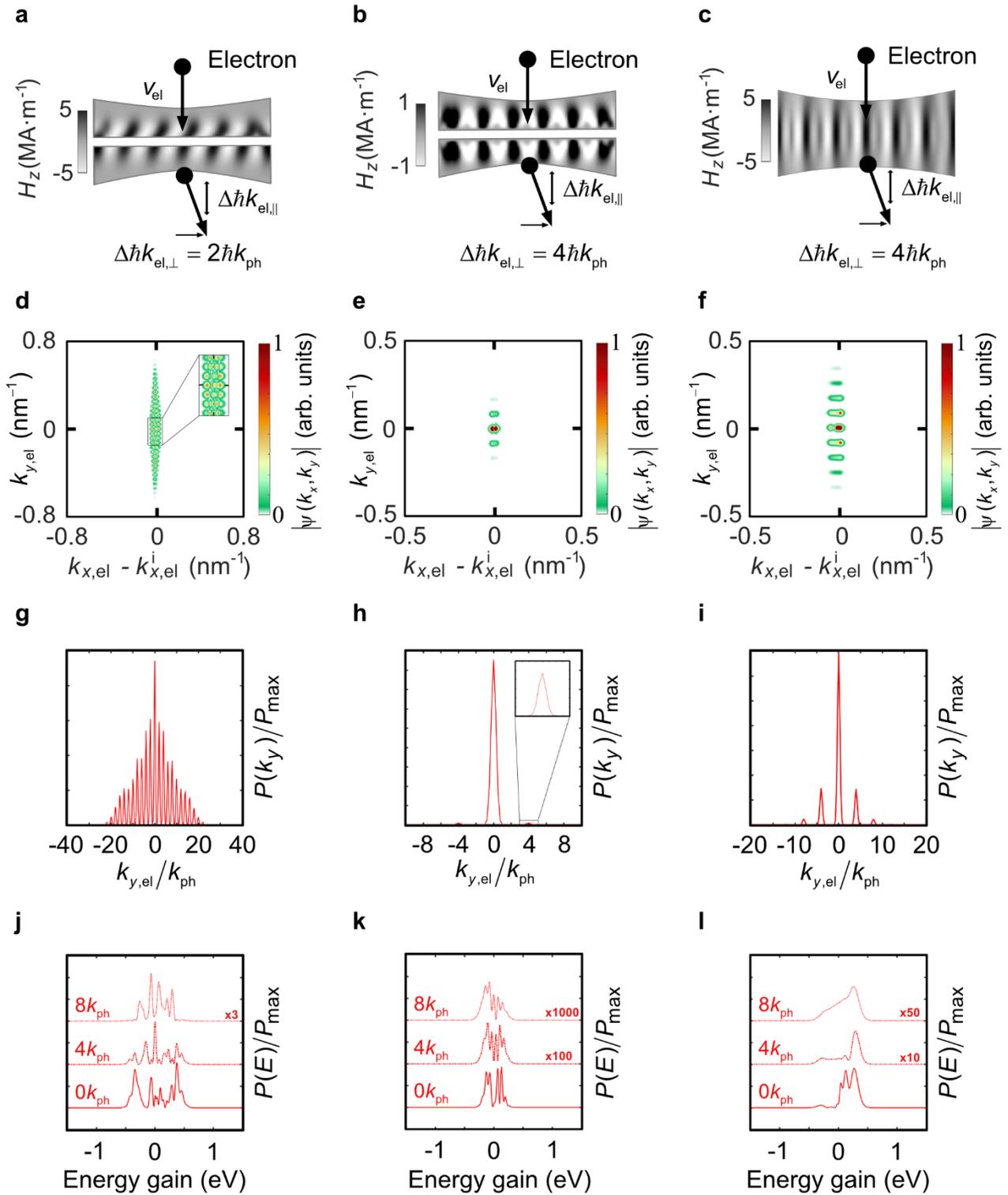

**Fig. 5. Transverse and longitudinal electron matter wave control.** Elastic and inelastic scattering of an electron wavepacket at counterpropagating pulsed Hermite-Gaussian (HG) beams of different spatial modes and wavelength. The considered superimposed spatial modes are $HG_{10}$ and $HG_{00}$ of $\lambda = 300$ nm (left; **a, d, g, j**), $HG_{10}$ at $\lambda = 300$ nm and $HG_{10}$ at $\lambda = 150$ nm (middle; **b, e, h, k**) and , $HG_{00}$ at $\lambda = 300$ nm and $HG_{00}$ at $\lambda = 150$ nm (right; **c, f, i, l**). For the numerical calculations all considered beams have a temporal full width

at half maximum (FWHM) of 10 fs. The electric-field-amplitudes are 5 GVm$^{-1}$ (HG$_{00}$) and 4.3 GVm$^{-1}$ (HG$_{10}$). The electron wavepacket has an initial kinetic energy of 1 keV, and initial longitudinal and transverse FWHM of $W_\mathrm{L} = 250$ nm and $W_\mathrm{T} = 60$ nm, respectively. **a-c**, Illustration of the scattering mechanism and depiction of the respective field distribution. **d-f**, Final transverse and longitudinal momentum population obtained from the three different numerical calculations. **g-i**, Final transverse momentum spectra in orders of photon momentum (all three consider $\lambda = 300$ nm). **j-l**, Final electron energy gain spectra depicted for three different diffraction orders for each of the three different numerical calculations. In **g-l**, the momentum distributions and electron energy gain spectra are normalized to the maximum observed probability amplitude and placed with an offset relative to each other. The inset in **j-l** are denoting the scaling factor for better visualization.

**Conclusion**

We discuss the Kapitza-Dirac effect as well as stimulated Compton scattering introduced by pulsed light beams that are spatially structured in higher order Gaussian beam profiles. This combination of short pulse duration and the transverse momentum distribution of these higher order Gaussian beams can satisfy the criteria of inelastic stimulated Compton scattering and result into the formation of a final electron energy-gain spectrum demonstrating PINEM-like distinct energy side bands. Simultaneously the electron wavepacket can be diffracted in distinct momentum orders where each momentum order demonstrate a modulated electron energy-gain spectrum as well.

To describe this interaction as completely as possible, we discussed transverse and longitudinal manipulation processes in terms of structuring function and pulse temporal duration separately. This led to the conclusion that structured light itself does not influence the diffraction orders of the Kapitza-Dirac effect. However, when combined with a finite pulse duration, it becomes possible to manipulate the electron diffraction pattern by controlling the interaction time and field amplitude through the appropriate choice of pulse duration and spatial mode. This leads for short and weaker interactions respectively to the conventional Kapitza-Dirac diffraction pattern. Contrary, for longer or stronger interactions it is possible to reach the strong interaction regime, characterized by the predominant population of higher diffraction orders and the influence of quantum-path interferences on the final electron momentum spectrum. By further varying the wavelength it is possible to tune the separation between the diffraction orders by $2k_\mathrm{ph}$ and $4k_\mathrm{ph}$. At the same time the electron energy spectrum can be coherently manipulated by carefully choosing the pulse duration and electron velocity. By additionally varying the combination of different spatial modes and respective wavelengths, the spectral positions of observed energy sidebands are tuned and furthermore influence the population of different energy orders. These combinations however come alongside the tradeoff of a reduced field intensity that the electron experiences. To achieve a combined full control of both energy spectrum and transverse momentum state population with free-space light beams, it is required to find a balance between pulse durations, that favor the strong interaction in the Kapitza-Dirac effect and on the contrary weakens the interaction experienced by the inelastic stimulated Compton scattering. Nevertheless, the interaction strength in both regimes strongly depends on the electric field intensity. Our findings lead to a framework on how to simultaneously control electron energy and momentum states in free space by light.

To observe the proposed interactions here experimentally, it is first important to generate intense femtosecond pulsed and spatially-structured light beams, a capability already demonstrated in laboratories[42–44]. The use of optical cavities might further amplify the experimental interaction[7]. Furthermore, it is crucial to note that the calculations in this study are based on a highly-coherent electron wavepacket (see supplementary Note 5). The spatial wavepacket broadening (FWHM) directly correlates with both the electron energy spread and electron pulse duration. This assumption implies a temporally coherent electron pulse, reflecting the self-similarity of electrons within the pulse. Therefore, a successful experiment requires a short electron pulse with a high temporal coherence. Current state-of-the-art ultrafast scanning electron microscopy (USEM) setups based on photoemission tips approach the limits of these parameters by providing electron pulses with suitable energy spread and pulse durations of around 200 fs[45,46]. Despite this, these pulses exhibit temporal coherence lengths spanning a few laser cycles and having demonstrated coherent electron-light interactions[37]. Further advancements in ultrashort electron pulse generation could be achieved by adopting innovative electron emission schemes such as cooled atoms[47], single-atom tip sources[48], single-crystalline surfaces[49], and superconducting niobium field emission tips[50]. These approaches hold promise for generating the necessary electron pulses for the envisioned experiment. The experimental implementation would therefore realize a free-space-based electron matter-wave control that holds the potential to enable new possibilities for momentum- and energy-resolved probing and excitations such as plasmon polaritons, plasmons, excitons and phonons[8]. Especially the ability of combining different spatial modes with each other could lead to the arbitrary design of the peak positions of the populated energy sidebands. This could be achieved by the combination of various spatial modes such as the full realm of Hermite-Gaussian beams or the combination with different spatially structured light fields such as Laguerre-Gaussian beams, Ince Gaussian beams and Bessel beams[51].

## Methods

### Power normalized higher-order Gaussian beams

The Hermite-Gaussian modes are derived from the paraxial Helmholtz equation for amplitudes that depend on the transverse coordinates $x$ and $z$ and are given by:

$$\vec{A}_{n,m}(x,y,z) = \vec{e}_y A_0 \frac{w_0}{W(y)} H_n\left(\frac{\sqrt{2}x}{W(y)}\right) H_m\left(\frac{\sqrt{2}z}{W(y)}\right) \exp\left(-\frac{x^2+z^2}{W(y)^2}\right)$$
$$\times \exp\left(i\left[|\vec{k}|y - (1+n+m)\arctan\frac{y}{y_0} + \frac{|\vec{k}|(x^2+z^2)}{2R(y)}\right]\right) \quad . \quad (6)$$

Consequently, these higher modes are described by a Gaussian envelope that is modulated by Hermite polynomials. This modulation results in $n$ and $m$ nodes in the transverse intensity distribution. In this work we reduce the complexity to a two-dimensional problem within the x-y plane. The outlined solution in in equation (1) is normalized to their field amplitude. This has the consequence that, for example, with varying beam waist or higher-order solutions, the solutions all exhibit the same maximum amplitude (see supplementary note 1). However, this is not the case when considering these solutions in the context of an experimental implementation. In experiments, the power of the laser beam is usually fixed. By varying the beams structural parameter (e.g. beam waist or structural order (n,m)) the power stays constant and beam amplitude varies accordingly. For a quantitative comparison between these solutions, one therefore should determine the power-normalized forms of these solutions. The normalization to unit power flow is achieved by considering the integral over the beam cross-section to be unity $\int_{-\infty}^{\infty} dx |\vec{A}(\vec{r},t)|^2 = 1$.

By applying this to the HG$_{10}$ and HG$_{00}$ modes in two dimensions, given by equation (6), one obtains the power normalized amplitude of the vector potential:

$$\vec{A}_{1,0}(x,y) = \vec{e}_y \frac{A_0}{(2\pi)^{1/4} \sqrt{w_0}} \frac{w_0}{W(y)} \frac{2\sqrt{2}x}{W(y)} \exp\left(-\frac{x^2}{W(y)^2}\right)$$
$$\times \exp\left(i\left[|\vec{k}|y - 2\arctan\frac{y}{y_0} + \frac{|\vec{k}|x^2}{2R(y)}\right]\right) \quad , \qquad (7)$$

$$\vec{A}_{0,0}(x,y) = \vec{e}_y \frac{A_0}{\sqrt{w_0}} \left(\frac{2}{\pi}\right)^{1/4} \frac{w_0}{W(y)} \exp\left(-\frac{x^2}{W(y)^2}\right)$$
$$\times \exp\left(i\left[|\vec{k}|y - \arctan\frac{y}{y_0} + \frac{|\vec{k}|x^2}{2R(y)}\right]\right) \quad . \qquad (8)$$

Consequently, the peak amplitude of the HG$_{10}$ and HG$_{00}$ modes will differ from each other. In this work we give the considered field amplitude as one of the essential numerical parameters of our calculations. Thereby the Gaussian field amplitude is considered as the maximum amplitude and as the reference. The peak amplitude of the HG$_{10}$, thus can be calculated by the amplitude from the power normalized expression at the position of the maximum at $x = w_0/\sqrt{2}$. Therefore, we obtain an amplitude given by $A_{1,0} = \sqrt{2/e}\, A_{0,0}$, where $e$ is the Euler number.

**Transverse momentum distribution and momentum distribution convolution**

The transverse momentum distribution of the considered Hermite-Gaussian spatial modes can be obtained by performing a Fourier transformation of the modes amplitude distribution[34]:

$$A_{1,0}(k_\perp) = \frac{1}{\sqrt{2\pi}} \int A_{1,0}(x, y=0) e^{-ik_\perp x} dx = A_0 w_0^{3/2} \frac{ik_\perp}{2(2\pi)^{1/4}} \exp(-k_\perp^2 w_0^2/4), \qquad (9)$$

$$A_{0,0}(k_\perp) = \frac{1}{\sqrt{2\pi}} \int A_{0,0}(x, y=0) e^{-ik_\perp x} dx = \frac{A_0 w_0^{3/2}}{(2\pi)^{1/4}} \exp(-k_\perp^2 w_0^2/4). \qquad (10)$$

To obtain the probability distributions of the super positioned spatial beam we perform a convolution of the superimposed probability distributions to obtain a new weighted probability distribution. For the superposition of the HG$_{10}$ and HG$_{00}$ spatial mode we obtain:

$$A_C(k_\perp) = A_{1,0}(k_\perp) * A_{0,0}(k_\perp) = \frac{A_0^2 w_{0,1}^{3/2} w_{0,2}^{3/2}}{2\sqrt{2\pi}} \int_{-\infty}^{\infty} k_{\perp,0} \exp\left(-k_{\perp,0}^2 w_{0,1}^2/4\right) \exp\left(-(k_\perp - k_{\perp,0})^2 w_{0,2}^2/4\right) dk_{\perp,0}$$

$$= \frac{A_0^2}{\sqrt{2}} \frac{w_{0,1}^{3/2} w_{0,2}^3}{(w_{0,1}^2 + w_{0,2}^2)^{3/2}} k_\perp \exp\left(-\frac{w_{0,1}^2 w_{0,2}^2 k_\perp^2}{4(w_{0,1}^2 + w_{0,2}^2)}\right)$$

(11)

This solution is also applicable to the superposition of two Hermite-Gaussian modes with the wavelength relation $\lambda_1 = 2\lambda_2$. In this case it is reasonable to approximate the transverse momentum probability distribution of the HG$_{10}$ beam at $\lambda_2$ as that of a HG$_{00}$ beam. For the convolution of two HG$_{00}$ spatial modes we obtain:

$$A_C(k_\perp) = A_{0,0}(k_\perp) * A_{0,0}(k_\perp) = \frac{A_0^2 w_{0,1}^{3/2} w_{0,2}^{3/2}}{\sqrt{2\pi}} \int_{-\infty}^{\infty} \exp\left(-k_{\perp,0}^2 w_{0,1}^2/4\right) \exp\left(-(k_\perp - k_{\perp,0})^2 w_{0,1}^2/4\right) dk_{\perp,0}$$

$$= A_0^2 \frac{\sqrt{2} w_{0,1}^{3/2} w_{0,2}^{3/2}}{(w_{0,1}^2 + w_{0,2}^2)^{1/2}} \exp\left(-\frac{w_{0,1}^2 w_{0,2}^2 k_\perp^2}{4(w_{0,1}^2 + w_{0,2}^2)}\right)$$

. (12)

For the convolution of two HG$_{10}$ spatial modes we obtain:

$$A_C(k_\perp) = A_{1,0}(k_\perp) * A_{1,0}(k_\perp) = \frac{A_0^2 w_{0,1}^{3/2} w_{0,2}^{3/2}}{4\sqrt{2\pi}} \int_{-\infty}^{\infty} k_{\perp,0}(k_\perp - k_{\perp,0}) \exp\left(-\frac{k_{\perp,0}^2 w_{0,1}^2}{4}\right) \exp\left(\frac{-(k_\perp - k_{\perp,0})^2 w_{0,1}^2}{4}\right) dk_{\perp,0}$$

$$= \frac{A_0^2}{16} \frac{w_{0,1}^{3/2} w_{0,2}^{3/2}}{(w_{0,1}^2 + w_{0,2}^2)^{5/2}} \left(w_{0,1}^2 w_{0,2}^2 k_\perp^2 - 2w_{0,2}^2 - 2w_{0,2}^2\right) \exp\left(-\frac{w_{0,1}^2 w_{0,2}^2 k_\perp^2}{4(w_{0,1}^2 + w_{0,2}^2)}\right)$$

(13)

**Kapitza-Dirac effect for a spatially structured standing-wave light: Analytical description**

We insert the power normalized amplitudes into the description of a standing-wave light. Doing that we obtain the following expressions for the vector-potential:

$$\vec{A}_{1,0}(\vec{r},t) = \frac{A_0}{(2\pi)^{1/4} \sqrt{w_0}} \frac{2\sqrt{2}}{w_0} x \exp\left(-\frac{x^2}{w_0^2}\right) \cdot \cos(\omega t) \cos\left(|\vec{k}|y\right) \hat{y},$$

(14)

$$\vec{A}_{0,0}(\vec{r},t) = \frac{A_0}{\sqrt{w_0}} \left(\frac{2}{\pi}\right)^{1/4} \exp\left(-\frac{x^2}{w_0^2}\right) \cdot \cos(\omega t) \cos\left(|\vec{k}|y\right) \hat{y}. \qquad (15)$$

These results can be inserted into the Volkov wave function for ponderomotive interactions[52]:

$$\psi_W(\vec{r},t) = \frac{1}{(2\pi)^{2/3}} \exp\left(i\vec{k}_{el} \cdot \vec{r} - i\omega_{el} t\right) \exp\left(-i\frac{e^2}{2\hbar m_0} \int_{-\infty}^{t} d\tau |A(\tau)|^2\right). \qquad (16)$$

In the following the calculation will be exemplary presented for the Gaussian-beam standing-wave light. For this it is necessary to calculate the squared vectorpotential $|A(\vec{r},t)|^2 = A_0^2/w_0 (2/\pi)^{1/2} \exp(-2x^2/w_0^2)\cos^2(\omega t)\cos^2(|\vec{k}|y)$. Furthermore, the temporal oscillation of $A_0^2 \cos^2(\omega t)$, that results in highly oscillatory phase variations at twice the light frequency, can be replaced with its average value $A_0^2/2$. Additionally, by rewriting the phase modulation in transverse direction $\cos^2(|\vec{k}|y) = 1/2(1+\cos(2|\vec{k}|y))$ it becomes possible to restructure the term to $|A(\vec{r},t)|^2 = A_0^2/(4w_0)(2/\pi)^{1/2}\exp(-2x^2/w_0^2)(1+\cos(2|\vec{k}|y))$. Inserting this into the ponderomotive phase argument and relating the propagation length $x$ to the delay $t'$ with $x = v_{el} t'$, the Volkov phase recast into:

$$\begin{aligned}\phi_W &= -\frac{e^2}{2\hbar m_0}\int_{-\infty}^{t} dt'|A(t')|^2 = -\frac{e^2}{2\hbar m_0 v_{el}}\int_{-\infty}^{\infty} dx|A(x)|^2 \\ &= -\frac{e^2}{2\hbar m_0 v_{el}}\frac{A_0^2}{4w_0}(2/\pi)^{1/2}\int_{-\infty}^{\infty} dx\exp(-2x^2/w_0^2)dx \cdot (1+\cos(2|\vec{k}|y)) \\ &= -\frac{e^2 A_0^2}{8\hbar m_0 v_{el}}(1+\cos(2|\vec{k}|y)).\end{aligned} \quad (17)$$

With this phase expression the Volkov wave function, while considering the Jakobi-Anger transformation $e^{ia\cos\theta} \equiv \sum_m i^m J_m(a) e^{im\theta}$, yields to:

$$\begin{aligned}\psi_W(\vec{r},t) &= \frac{1}{(2\pi)^{2/3}}\exp(ik_{el}x - i\omega_{el}t)\exp\left(-i\frac{e^2 E_0^2}{8\hbar m_0 v_{el}\omega^2}\right) \\ &\times \sum_m i^m J_m\left(\frac{e^2 E_0^2}{8\hbar m_0 v_{el}\omega^2}\right)\exp(i2m|\vec{k}|y)\end{aligned}. \quad (18)$$

Here, the electric field amplitude was introduced with $E_0 = \omega A_0$. The probability to populate the $m^{th}$ diffraction order is therefore given by:

$$P_m^{HG_{00}} = J_m^2\left(\frac{e^2 E_0^2}{8\hbar m_0 v_{el}\omega^2}\right). \quad (19)$$

When repeating this calculation for the HG$_{10}$ standing light wave, the exact same result is obtained.

**Kapitza-Dirac effect for a pulsed standing light wave: Analytical description**

The vector potential for a pulsed standing-wave is given by:

$$\vec{A}_\tau(\vec{r},t) = A_0 \exp\left(-\frac{t^2}{4\tau^2}\right)\cdot \cos(\omega t)\cos(|\vec{k}|y)\hat{y}. \quad (20)$$

This field can also be inserted separately into the Volkov wave function for ponderomotive interactions (equation 26). The necessary squared vector potential is given by:

$$\left|\vec{A}_\tau(\vec{r},t)\right|^2 = A_0^2 \exp\left(-\frac{t^2}{2\Delta\tau^2}\right) \cdot \cos^2(\omega t)\cos^2\left(\left|\vec{k}\right|y\right). \tag{21}$$

This expression again can be simplified by approximating $A_0^2 \cos^2(\omega t) \approx A_0^2/2$ and rewriting the transverse phase modulation $\cos^2\left(\left|\vec{k}\right|y\right) = 1/2\left(1+\cos\left(2\left|\vec{k}\right|y\right)\right)$ so that we obtain $\left|\vec{A}_\tau(\vec{r},t)\right|^2 = A_0^2 \exp\left(-t^2/2\Delta\tau^2\right)\cdot\cos^2(\omega t)\cos^2\left(\left|\vec{k}\right|y\right)$. Inserting this expression into the ponderomotive phase argument, while relating the propagation length $x$ to the delay $t'$ with $x = v_{el}t'$, yields:

$$\phi_W = -\frac{e^2}{2\hbar m_0}\int_{-\infty}^{t} dt'\left|A(t')\right|^2 = -\frac{e^2}{2\hbar m_0 v_{el}}\int_{-\infty}^{\infty} dx\left|A(x)\right|^2$$

$$= -\frac{e^2}{2\hbar m_0 v_{el}}\frac{A_0^2}{2}\int_{-\infty}^{0} dx\exp\left(-x^2/2(v_{el}\Delta\tau)^2\right)dx\cdot\left(1+\cos(2\left|\vec{k}\right|y)\right). \tag{22}$$

$$= -\sqrt{\frac{\pi}{2}}\frac{e^2 A_0^2}{\hbar m_0}\Delta\tau\left(1+\cos(2\left|\vec{k}\right|y)\right).$$

With this phase expression the volkov wave function, while considering the Jakobi-Anger transformation $e^{ia\cos\theta} \equiv \sum_m i^m J_m(a)e^{im\theta}$, yields to:

$$\psi_W(\vec{r},t) = \frac{1}{(2\pi)^{2/3}}\exp\left(ik_{el}x - i\omega_{el}t\right)\exp\left(-i\sqrt{\frac{\pi}{2}}\frac{e^2 E_0^2}{\hbar m_0 \omega^2}\Delta\tau\right)$$
$$\times \sum_m i^m J_m\left(\sqrt{\frac{\pi}{2}}\frac{e^2 E_0^2}{\hbar m_0 \omega^2}\Delta\tau\right)\exp\left(i2m\left|\vec{k}\right|y\right) \tag{23}$$

Here, the electric field amplitude was introduced with $E_0 = \omega A_0$. The probability to populate the $m^{th}$ diffraction order is therefore given by:

$$P_m^\tau = J_m^2\left(\sqrt{\frac{\pi}{2}}\frac{e^2 E_0^2}{\hbar m_0 \omega^2}\Delta\tau\right). \tag{24}$$

This solution reveals that the scattering order determined by the Bessel functions and further demonstrates that the electron wave function after the interaction can be seen as a combination of plane wave electrons occupying (m, -m) momentum states.

**Time-dependent Maxwell-Schrödinger algorithm**[25,35,53]

To simulate the dynamics of the electron wavepacket, we employed the minimal-coupling Hamiltonian, which encompasses the temporally and spatially varying vector potential. The formulation of the time-dependent Schrödinger equation, incorporating the minimal-coupling Hamiltonian, is as follows:

$$i\hbar \frac{\partial}{\partial t}\psi(\vec{r},t) = \left[ -\frac{\hbar^2}{2m_0}\nabla^2 + \frac{e^2}{2m_0}\left|\vec{A}(\vec{r},t)\right|^2 - \frac{i\hbar e}{m_0}\vec{A}(\vec{r},t)\cdot\vec{\nabla} \right]\psi(\vec{r},t). \qquad (25)$$

Here, we applied the coulomb gauge $\vec{\nabla}\cdot\vec{A}=0$ and $\psi(\vec{r},t)$ is the time-dependent electron wave function. Further, $\vec{A}(\vec{r},t)$ is the vectorpotential, $m_0$ is the electron mass, $\hbar$ is the reduced Planck's constant, and $e$ is the electron charge. The time propagator is approximated using a second-order differencing scheme, while spatial differentiation is accomplished through a series of steps, starting with a Fourier transformation, continuing with the multiplication by a suitable transfer function, and lastly the inverse Fourier transformation. Commonly known as the Fourier method, this approach provides stability and faster convergence in comparison to finite differentiation in molecular dynamics simulations[35,54]. Furthermore, the accuracy of the results is confirmed by ensuring that the norm of the wave function remains closely aligned with $N = \int d^3r |\psi(\vec{r},t)|$ at each given time. The overall longitudinal momentum distribution is given by an integration of the form $P(k_y) = \int dk_x |\tilde{\psi}(k_x,k_y)|$ for each time step. This can be related with energy momentum relation $E = \hbar^2 k_y^2/(2m_0)$ to the overall electron energy-gain spectrum $\Sigma(E)$. The overall transverse momentum distribution is given by an integration of the form $P(k_x) = \int dk_y |\tilde{\psi}(k_x,k_y)|$ for each time step. The initial electron state is modeled by an analytical expression for a Gaussian wavepacket[53]:

$$\psi_0(x,y,t=0) = \exp(ik_{x,\text{el}}^i x)\left(\frac{1}{2\pi W_T W_L}\exp\left\{-\frac{1}{2}\frac{(x-x_0)^2}{W_L^2}\right\}\exp\left\{-\frac{1}{2}\frac{(y-y_0)^2}{W_T^2}\right\}\right)^{\frac{1}{2}}. \qquad (26)$$

For this $W_L$ and $W_T$ define the longitudinal and transverse broadening FWHM of the electron wavepacket. The wavepacket is initially centered around $(x_0, y_0)$ and has the initial electron momentum of $k_{x,el}^{(i)} = m_0 v_{el}/\hbar$. To calculate the numerical solution for Maxwell's equations, we utilized a finite difference time-domain method implemented in a MATLAB environment through a custom-written numerical code[55]. Thus, the electromagnetic fields are obtained for each time step and related to the vectorpotential by $\vec{B} = \vec{\nabla}\times\vec{A}$ and $\vec{E} = -\partial_t \vec{A}$. The vector potentials originating from the Maxwell simulation domain were subsequently mapped onto the Schrödinger simulation domain. Both domains were discretized using piecewise linear unit cells, with the flexibility for varying sizes of discretization units. For the calculations in this work, we used $\delta x = \delta y = 3\,nm$ for the Maxwell domain, and $\delta x = \delta y = 1.8\,nm$ for the Schrödinger domain. Based on our experience with this numerical toolbox, the main limiting factor that influences the achievement of numerically convergent results is the Courant number of the finite-difference time-domain method ($S_c = c\delta t/\delta x$) and the relative size of the unit cells in the simulation domains. Therefore, achieving a balance between accuracy and convergence is crucial to maintain reasonable simulation times.

**Data availability**

Additional data supporting the conclusions presented in both the main text and Supplementary Information can be obtained from the corresponding authors upon a reasonable request.

**Code availability**

The computer codes underpinning the results outlined in both the main text and Supplementary Information can be obtained from the corresponding authors upon a reasonable request.

**Acknowledgement**

We thank C. Wolff for stimulating discussions. This project has received funding from the European Research Council (ERC) under the European Union's Horizon 2020 research and innovation program under grant agreement no. 802130 (Kiel, NanoBeam) and grant agreement no. 101017720 (EBEAM). Sven Ebel acknowledges support by the Danish National Research Foundation (Project No. DNRF165).

**Author contributions**

N. T. initiated and supervised the project. S. E. conceived the idea and carried out simulations, analytical work and data analysis. S.E. and N.T. wrote the manuscript.

**Competing interests**

The authors declare no competing interests.

# Structured free-space optical fields for transverse and longitudinal control of electron matter waves supplementary information

**Sven Ebel[1, *] and Nahid Talebi[2,3, ⊥]**

[1]POLIMA—Center for Polariton-driven Light-Matter Interactions, University of Southern Denmark, 5230 Odense, Denmark

[2]Institute of Experimental and Applied Physics, Kiel University, 24098 Kiel, Germany

[3]Kiel Nano, Surface and Interface Science KiNSIS, Kiel University, 24118 Kiel, Germany

⊥ talebi@physik.uni-kiel.de, * sleb@mci.sdu.dk

**Supplementary contents**

Supplementary Note 1: Power normalization and amplitude normalization

Supplementary Note 2: Inelastic stimulated Compton scattering

Supplementary Note 3: Kapitza-Dirac effect for a spatially structured standing-wave light pattern – Bragg and Raman-Nath regime

Supplementary Note 4: Effective amplitude of a spatially structured pulsed light beam

Supplementary Note 5: Influence of the longitudinal broadening of the electron wavepacket

Supplementary Note 6: Kapitza-Dirac effect for a spatially structured pulsed standing-wave light - Interaction dynamics

Supplementary Note 7: Inelastic electron scattering from pulsed spatially structured traveling light waves

Supplementary Note 8: Influence of the optical pulse duration and pulse synchronization on the transversal momentum and final electron energy spectrum

Supplementary Note 9: Dis-entangling field intensity and pulse duration

Supplementary Note 10: Variation of the electron velocity

Supplementary Note 11: Interaction dynamics for different spatial modes

**Supplementary Note 1: Power normalization and amplitude normalization**

We illustrate the comparison between the amplitude (Figure S1a) and power normalized (Figure S1b) Gaussian beam for varying beam waists. Both solutions have been introduced in the main manuscript. However, all the numerical calculation presented in this work are obtained for the power normalized solution.

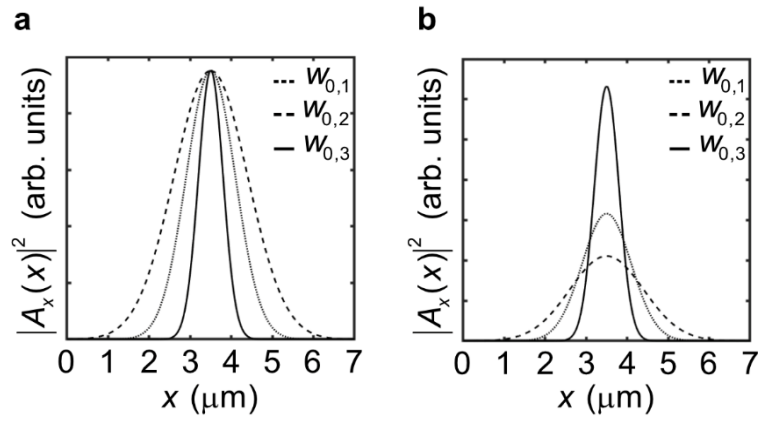

**Figure S1** Amplitude normalized (a) and power normalized (b) beam amplitude profiles for varied beam waists $w_{0,1} = 6\lambda$, $w_{0,2} = 4\lambda$, and $w_{0,3} = 2\lambda$ of a Gaussian beam. The considered wavelength is $\lambda = 300$ nm.

**Supplementary Note 2: Inelastic stimulated Compton scattering**

The broad transversal momentum spectrum of spatially structured light beams can satisfy the criterion of stimulated Compton scattering when the considered beams are pulsed in time. Here, we depict the various transversal momentum distributions of $HG_{00}$ and $HG_{10}$ beams and the respective transverse momentum distributions obtained by convoluting the superimposed distributions with various spatial modes and wavelength (Figure S2a). Additionally, we illustrate the condition of inelastic stimulated Compton scattering for a $HG_{10}$ beam at $k_\perp^{max}$ versus wavelength and pulse duration $\Delta\tau$ (Figure S2b).

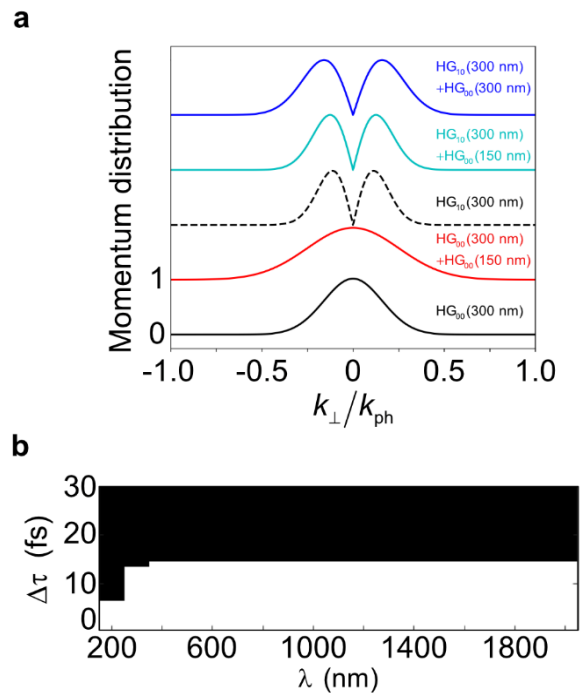

**Figure S2** (a) The effective transverse momentum distributions, obtained through convolution, of superimposed Hermite-Gaussian (HG) modes with varying orders and wavelengths, compared to the transverse momentum distributions of the fundamental $HG_{10}$ and $HG_{00}$ modes ($HG_{00}$(300nm)). The considered wavelength for the latter is $\lambda = 300$ nm. (b) Fulfillment of the inelastic stimulated Compton scattering condition (white=true; black=false) for two counterpropagating $HG_{10}$ beams and an electron velocity of 1 keV in dependence of the pulse duration and wavelength.

**Supplementary Note 3: Kapitza-Dirac effect for a spatially structured standing-wave light pattern – Bragg and Raman-Nath regime**

The Kapitza-Dirac effect has two interaction regimes; namely, the Bragg scattering regime, where the electron is diffracted into one single diffraction order, and the diffractive regime resulting in a symmetric diffraction into many momentum states [1]. The boundary between Bragg scattering and diffractive scattering can be derived considering the energy time uncertainty relation. The interaction time of the electron with the standing wave is given by $\Delta t = \Delta w / v_{el}$, $\Delta w$ is in this case the width of the laser beam. Therefore, the interaction energy becomes uncertain by $\Delta E = \hbar/(2\Delta t)$. If the beam is focused narrowly, the interaction time is very short and consequently $\Delta E$ is much larger than the recoil shift $\delta_D = (2\hbar k_{ph})^2/(2m_0)$, causing the electron to diffract into many diffraction orders. The Bragg regime is observed for wider laser beams. In this situation, the interaction time of the electron with the light field is long and $\Delta E$ is much smaller than the electron's recoil shift $\delta_D$. In this case the electron cannot interact with two photons unless a special incident angle is chosen. This condition is satisfied at the Bragg angle $\theta_{Bragg}$. From this discussion it is possible to find a quantitative criterion $\rho = \delta_D/\Delta E$ for distinguishing these two regimes [1]. For $\rho \ll 1$ the diffraction regime dominates while for $\rho \gg 1$ the Bragg regime is dominating. It is now possible to discuss this criterion in terms of the spatial extent of the laser field. The energy uncertainty can be restructured as following: $\Delta E = \hbar v_{el}/(4w_0)$ while assuming an electron traveling distance of $2w_0$.

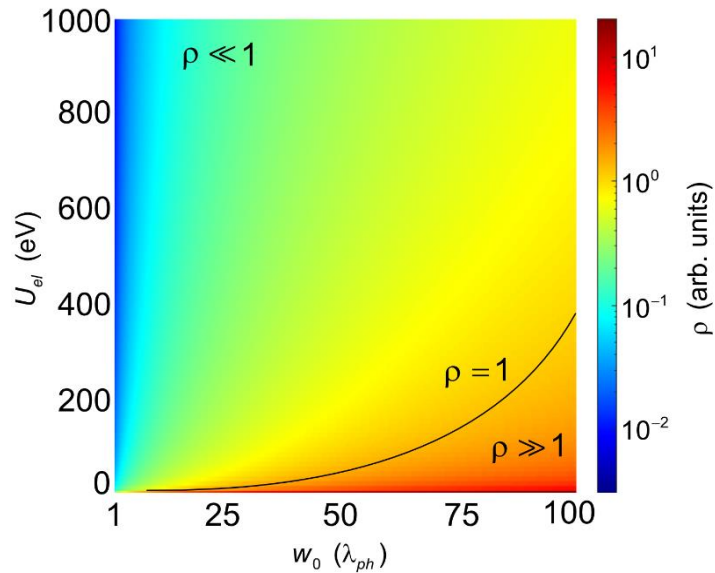

**Figure S3** Transition between the diffraction (Raman-Nath) and Bragg regimes. The figure shows the logarithmic distribution of ρ as a function of beam waist (in units of λ) and initial electron kinetic energy. The wavelength of the light is chosen as $\lambda = 300\ nm$. The black solid line marks $\rho = 1$.

This leads to the following criterion: $\rho = 4w_0 (2\hbar k_{ph})^2/(2m_0 \hbar v_{el})$ which is now a function of electron velocity and beam waist. This criterion demonstrates (see Figure S3) that for narrowly focused laser beams and electron velocities common in electron microscopes the electron will always experience the Kapitza-Dirac effect in the diffraction regime ($\rho \ll 1$).

**Supplementary Note 4: Effective amplitude of a spatially structured pulsed light beam**

The pulse duration strongly influences the strength of elastic scattering between the electrons and light beams. To illustrate the impact of choosing light beams with different spatial modes and a constant finite pulse duration, we illustrate the amplitude area that a point-like electron experiences during a 10-fs interactions with $HG_{10}$ and $HG_{00}$ spatial modes (see Figure S4), in the case of a perfect synchronization between the electron and the light pulse.

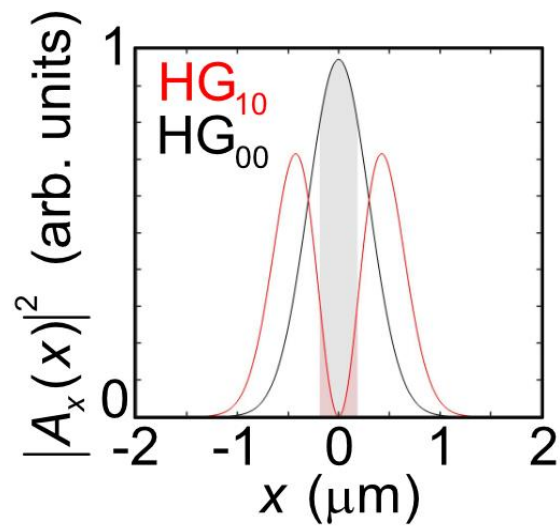

**Figure S4** Amplitude profiles of power normalized $HG_{10}$ (red) and $HG_{00}$ (black) distributions. The highlighting region is the associated area that a point like electron experience when interacting with these beams for a duration of 10 fs around the center.

**Supplementary Note 5: Influence of the longitudinal broadening of the electron wavepacket**

Here we compare the electron energy-gain spectra obtained for a Gaussian electron wavepacket having a varying longitudinal full width at half maximum (FWHM) broadening of $W_{L,1} = 50$ nm and $W_{L,2} = 250$ nm, and transversal FWHM broadening of $W_T = 60$ nm (see Fig S5). The electron wavepackets thereby interacts with two counterpropagating beams of the same HG spatial mode ($HG_{10}$) with central wavelength of $\lambda = 300$ nm, pulse duration of $\Delta\tau = 10$ fs, beam waist of $w_0 = 2\lambda$ and electric field amplitude of 8.6 GVm$^{-1}$. The electron has an initial center kinetic energy of 1 keV. For the small spatial extend of $W_{L,1} = 50$ nm, the electron energy gain spectrum does not demonstrate the formation of distinct sidebands. Furthermore, the spectrum is strongly asymmetric, by only undergoing an energy gain process.

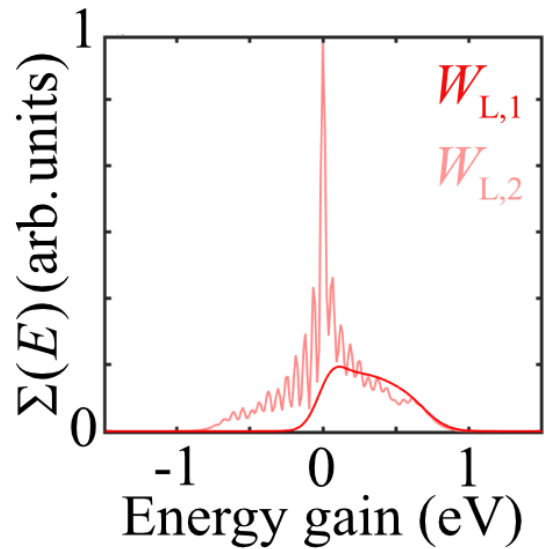

**Figure S5** Final electron energy-gain spectrum for an electron wavepacket with the spatial longitudinal broadening of $W_{L,1} = 50\ nm$ or $W_{L,2} = 250\ nm$ after the interaction with a $HG_{10}$ standing-wave light. The light pulse duration is 10 fs. The considered light fields have the electric-field-amplitude and wavelength of 8.6 GVm$^{-1}$ and 300 nm, respectively. The electron wavepacket has an initial kinetic energy of 1 keV.

**Supplementary Note 6: Kapitza-Dirac effect for a spatially structured pulsed standing-wave light – Interaction dynamics**

Here we provide the interaction dynamics obtained for the interaction of a Gaussian electron wavepacket, with $W_L = 250$ nm, $W_T = 60$ nm and the center kinetic energy of 1 keV with two counterpropagating beams of the same Hermite-Gaussian spatial mode, where we consider two cases with $HG_{10}$ (Figure S6) and $HG_{00}$, (Figure S7) distributions as the following. The beams have a central wavelength of $\lambda = 300$ nm, pulse duration of $\Delta\tau = 10$ fs, beam waist of $w_0 = 2\lambda$ and electric field amplitudes of 5 GVm$^{-1}$ ($HG_{00}$) and 4.3 GVm$^{-1}$ ($HG_{10}$), respectively. The dynamics obtained for the interaction with the $HG_{00}$ spatial mode (Figure S6) do not demonstrate a significant difference compared to the dynamics obtained for the $HG_{10}$ spatial mode (Figure S7). Additionally, we display the final electron energy gain spectrum within a few selected diffraction orders after the interaction with the $HG_{00}$ (black) and $HG_{10}$ (red) standing-wave light beams (Figure S8).

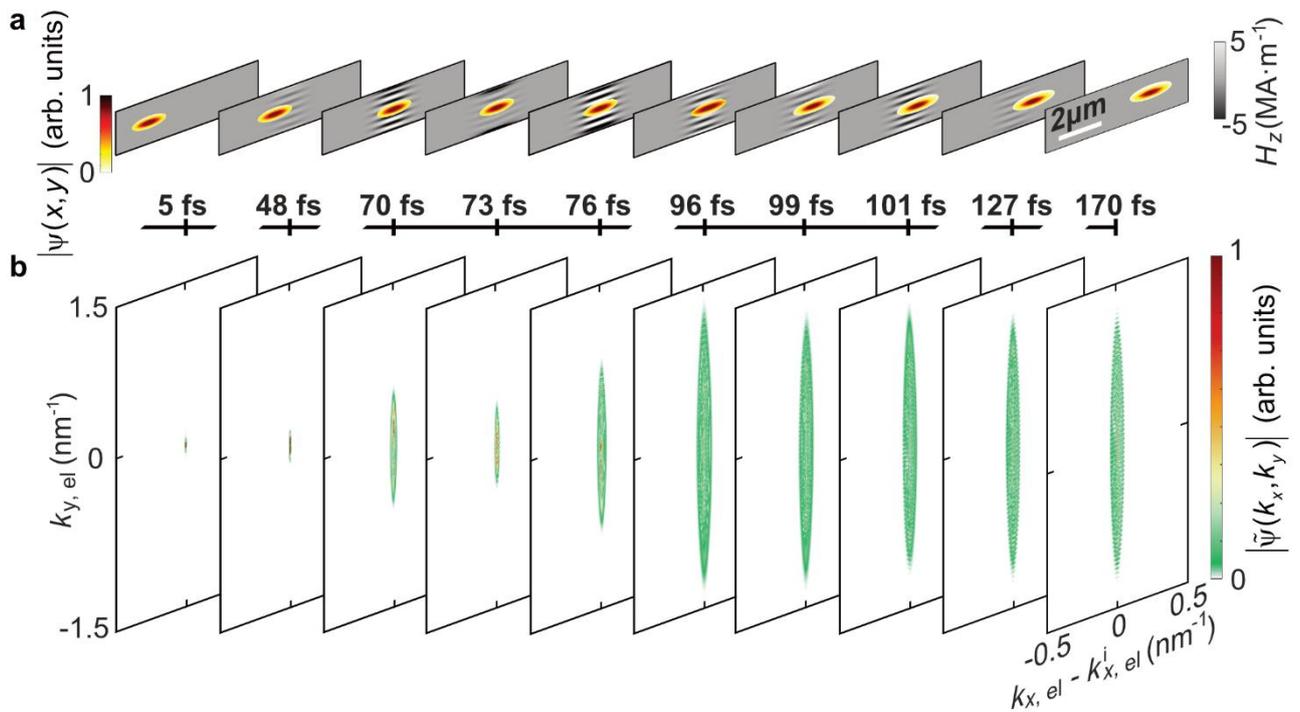

**Figure S6** Dynamics of the evolution of a Gaussian electron wavepacket through two counterpropagating Hermite-Gaussian ($HG_{00}$) pulsed laser beams (the laser electric-field-amplitude, its wavelength, and its temporal FWHM broadening are 5 GVm$^{-1}$, 300 nm and 10 fs, respectively) at different selected time steps. The electron wavepacket has an initial carrier energy of 1 keV and initial longitudinal and transverse FWHM of 250 nm and 60 nm, respectively. (a) The $z$-component of the magnetic field representing the formed structured light field (gray background) at depicted time steps, with the insets demonstrating the amplitude of the electron wavepacket. (b) Electron momentum distribution at the corresponding time steps.

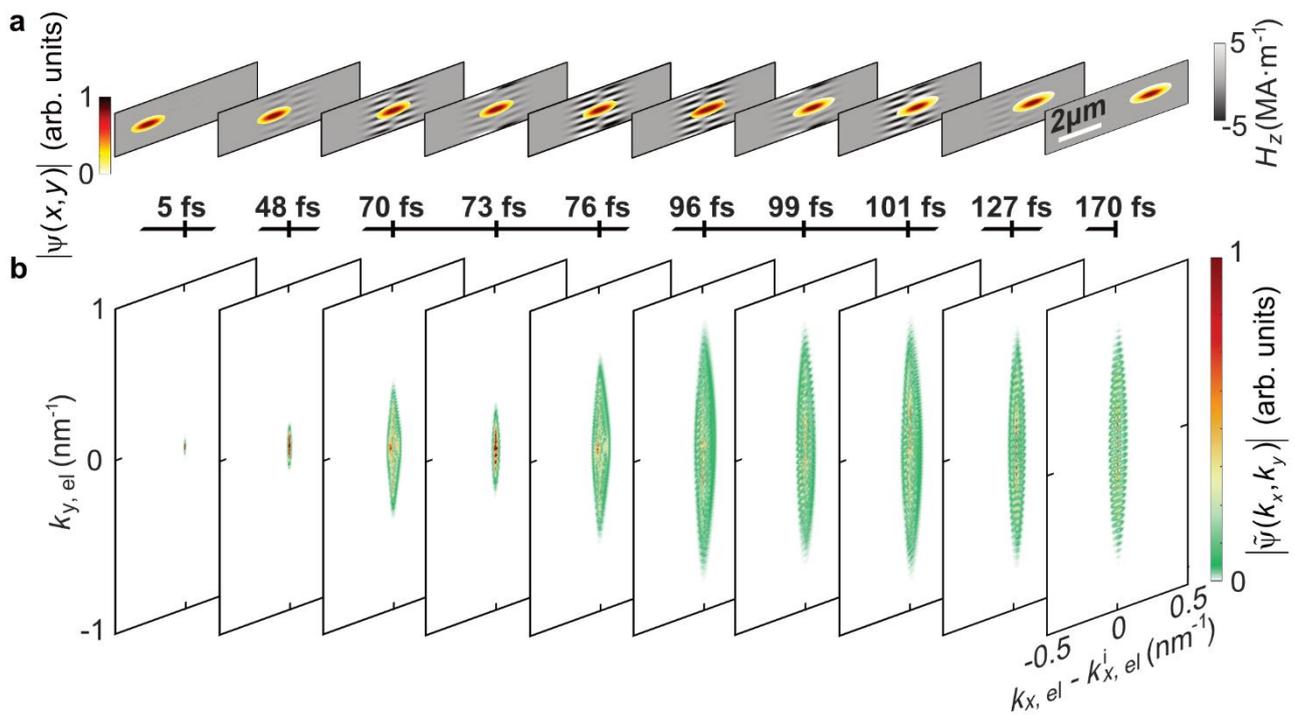

**Figure S7** Dynamics of the evolution of a Gaussian electron wavepacket through two counterpropagating Gaussian (HG$_{10}$) pulsed laser beams (the laser electric-field-amplitude, its wavelength, and its temporal FWHM broadening are 4.3 GVm$^{-1}$, 300 nm and 10 fs, respectively) at different selected time steps. The electron wavepacket has an initial carrier energy of 1 keV and initial longitudinal and transverse FWHM of 250 nm and 60 nm, respectively. (a) The z-component of the magnetic field representing the formed structured light field (gray background) at depicted time steps, with the insets demonstrating the amplitude of the electron wavepacket. (b) Electron momentum distribution at the corresponding time steps

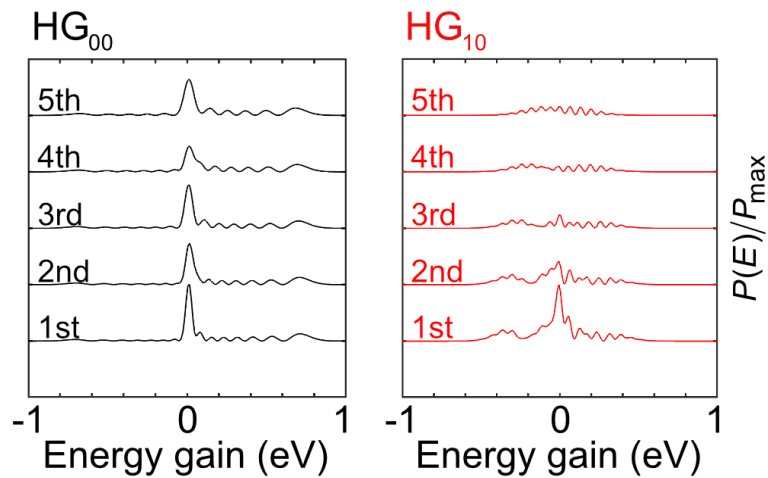

**Figure S8** Final energy gain spectra after the interaction with the HG$_{00}$ (black) and HG$_{10}$ (red) standing light waves. Displayed are the energy gain spectra obtained for the 1$^{st}$ to 5$^{th}$ diffraction order.

**Supplementary Note 7: Inelastic electron scattering from pulsed spatially structured traveling light waves**

To provide a full picture of the free space electron light interactions with spatially structured pulsed light fields we provide here a short numerical discussion of the interaction of an electron wavepacket with a single structured pulsed propagating beam. This interaction has been discussed in terms of a matter wave picture including the inelastic scattering from the ponderomotive potential formed by a traveling structured light pulsed wave [2]. The population of the distinct sidebands was explained in terms of electron wavepacket self-interference in the ponderomotive potential formed by the amplitude of the structured light wave. Here we revisit this interaction in terms of the interaction parameters discussed in the main manuscript. Therefore, a Gaussian electron wavepacket with $W_\mathrm{L} = 250$ nm, $W_\mathrm{T} = 60$ nm and the center kinetic energy of 1 keV interacts with a counterpropagating HG$_{10}$ or HG$_{00}$ beam. The respective beams have a central wavelength of $\lambda = 300$ nm, pulse duration of $\Delta\tau = 10$ fs, beam waist of $w_0 = 2\lambda$ and electric-field amplitudes of 5 GVm$^{-1}$ (HG$_{00}$) and 4.3 GVm$^{-1}$ (HG$_{10}$). The obtained interaction dynamics for the electron interacting with the traveling HG$_{10}$ beam reveals transversal momentum state population while showing an oscillatory behavior during the population of the longitudinal momentum states (Figure S9a). This oscillatory behavior occurs while the electron traverses the dual-node structure formed by the HG$_{10}$ beam. As a result, the final electron energy-gain spectrum demonstrates the formation of distinct sidebands (Figure S9c). The sideband distance is uniform and determined by $\Delta E = 59 \pm 1$ meV. This uniform sideband distance agrees very well to the energy gain expected from inelastic stimulated Compton scattering given for the HG$_{10}$ beam considered here by $\Delta E = 2\hbar v_\mathrm{el} k_\perp^{max} = 58$ meV. This leads to the conclusion that also for a single traveling pulsed HG$_{10}$ beam it is possible to fulfill the criteria of inelastic stimulated Compton scattering. This is explainable by considering the transverse momentum spectra of a single HG$_{10}$ beam (see supplementary note 2) which demonstrates two centrosymmetric maxima. Thus, the electron can interact with two photons, carrying a transverse photon momentum of $k_\perp^{max}$ and $-k_\perp^{max}$, respectively. This explanation represents a complementary explanation based on the particle viewpoint to the previously developed matter wave model [2]. Similarly, when considering the interaction with the traveling HG$_{00}$ beam the interaction dynamics demonstrate again instantaneous population occupations, occurring due to the non-energy-conserving interactions that take place within the non-rotating-wave picture, that vanishes again at the end of the interaction. Additionally, an oscillatory population of longitudinal momentum states is observable which we believe that caused by the back reflection of the rising and descending flanks of the Gaussian-shaped ponderomotive potential (Figure S9 c). This population, however, appears weaker in comparison to the interaction with the traveling HG$_{10}$ beam. Nevertheless, the final electron energy-gain spectrum demonstrates the formation of a few distinct sidebands with a separation of $\Delta E = 76 \pm 20$ meV. The energy separation and its corresponding error are larger than those obtained from the interaction with the traveling HG$_{10}$ beam. From the perspective of inelastic stimulated Compton scattering, this can be explained by the broad spectrum of transverse momentum values supported by a Gaussian beam that can satisfy the condition of stimulated Compton scattering. This leads to the conclusion that any spatial mode with a transverse momentum distribution, that can satisfy the criterium of inelastic stimulated Compton scattering, can cause the formation of distinct energy sidebands. Thus, we conclude that the overall strength of the traveling light beam and electron interactions should be still determined by the field intensity and electron velocity [2].

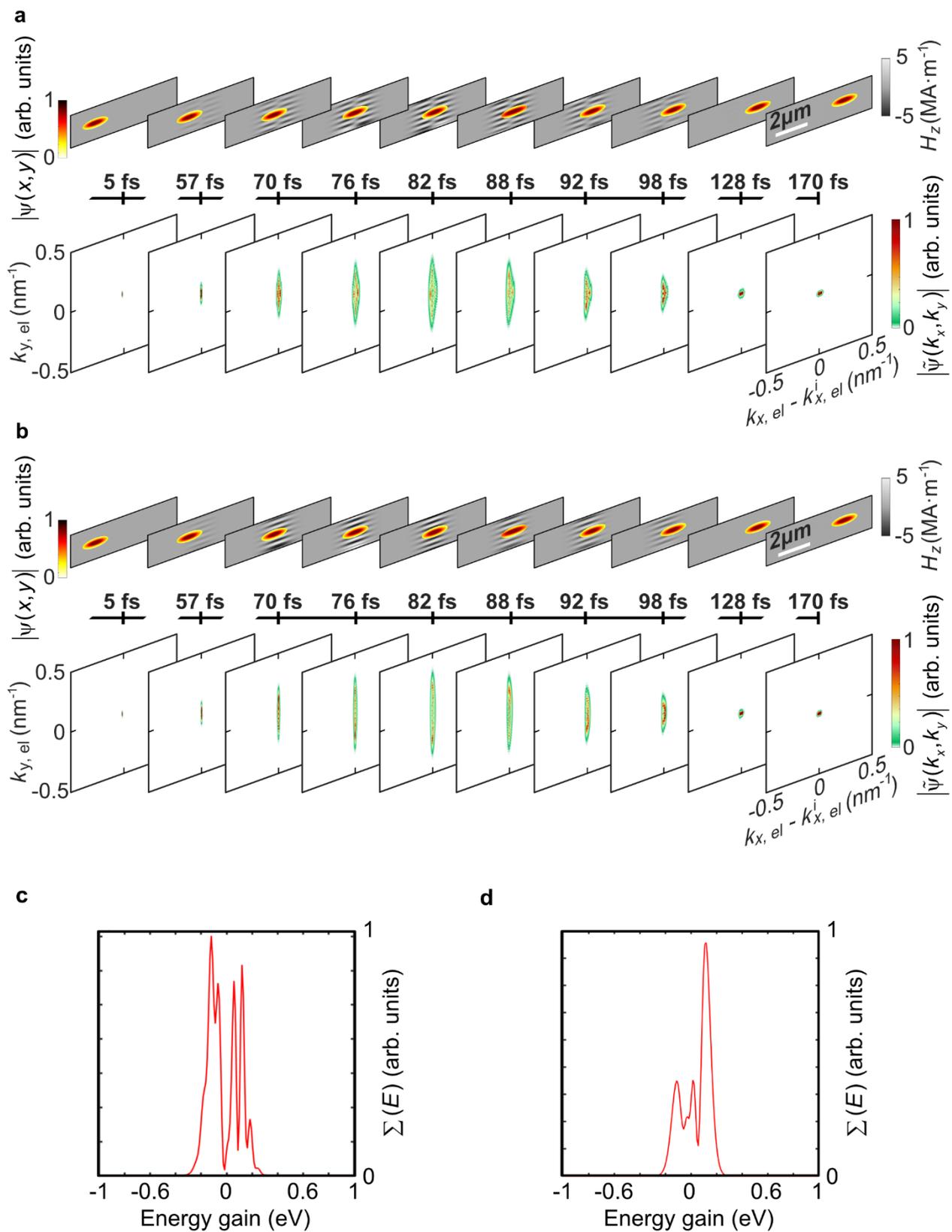

**Figure S9** Dynamics and final electron energy gain spectra obtained for the evolution of a Gaussian electron wavepacket in spatial and momentum space through a traveling first order Hermite-Gaussian (HG$_{10}$) (a, c) and Gaussian (HG$_{00}$) (b, d) beam (the laser electric-field-amplitude, its wavelength, and its temporal FWHM

broadening are 5 GVm$^{-1}$ (HG$_{00}$) and 4.3 GVm$^{-1}$ (HG$_{10}$), 300 nm and 10 fs, respectively). The electron wavepacket has an initial carrier energy of 1 keV and initial longitudinal and transverse FWHM of 250 nm and 60 nm, respectively. (a, b) The dynamics of the structured light field and the electron wavepacket are visualized by the *z*-component of the magnetic field (gray background). The inset demonstrating the amplitude of the electron wavepacket at different time steps. Furthermore, for each interaction, the obtained electron momentum distribution is visualized at the corresponding time steps.

**Supplementary Note 8: Influence of the optical pulse duration and pulse synchronization on the transversal momentum and final electron energy spectrum**

Here, we briefly discuss the influence of the optical pulse duration on both the transversal and longitudinal manipulation of the electron momentum distribution. Therefore, we vary for the two counterpropagating pulsed HG$_{10}$ beams their respective pulse duration in the combinations $\Delta\tau_1 = 20$ fs and $\Delta\tau_2 = 10$ fs, $\Delta\tau_1 = 10$ fs and $\Delta\tau_2 = 10$ fs, and finally $\Delta\tau_1 = 5$ fs and $\Delta\tau_2 = 10$ fs. The obtained results clearly demonstrate an impact on the number of occupied energy sidebands (see Figure S10 (a,c)). The transverse momentum spectrum reveals that, in the case of identical pulse duration, the electron scatters into a high number of diffraction orders. The momentum population exhibits clear indications of a strong interaction, with the highest diffraction probabilities observed for higher-order diffraction orders (identical with the results presented in the main manuscript). In the case where $\Delta\tau_1 = 2\Delta\tau_2$, the electron still undergoes diffraction into the same number of orders, however with different probability amplitudes for each order. Simultaneously, the energy-gain spectrum shows an overall weaker energy gain and loss probability amplitudes, with an increased population of the zero and first order energy sideband compared to the case of $\Delta\tau_1 = \Delta\tau_2$. The final scenario where $\Delta\tau_1 = 0.5\Delta\tau_2$, the overall interaction strength is weaker for both the transversal momentum and electron energy manipulation. This is illustrated by both a reduce number of populated momentum orders as well as the overall reduced extend of the electron energy gain spectrum. Additionally, the final transversal momentum spectrum does not show the characteristic of a strong interaction. We conclude with these results that the choice of the pulse duration is relevant when controlling the final electron momentum distribution. In particular, the case of an identical pulse duration for both beams one can expect the strongest possible interaction (for this specific parameter set).

The notable correlation between pulse duration and the strength of the electron-light interaction initiates an exploration of how this interaction is influenced by the synchronization between the arrival of the electron wavepacket and the light pulse. This influence is demonstrated by simulating the interaction at various delay times $t_0$ between the arrival of the electron wavepacket and the laser pulse. The arrival time hereby refers to the simultaneous arrival of both the center of the Gaussian electron wavepacket and the Gaussian pulse envelope at the center of the simulation domain. The synchronization dependency in Figure S10 (b, d) is characterized by a transition from the realm of strong interaction to a classical Kapitza-Dirac transverse momentum spectrum as the delays are extended. Additionally, one observes an asymmetric electron energy spectrum for longer delay times.

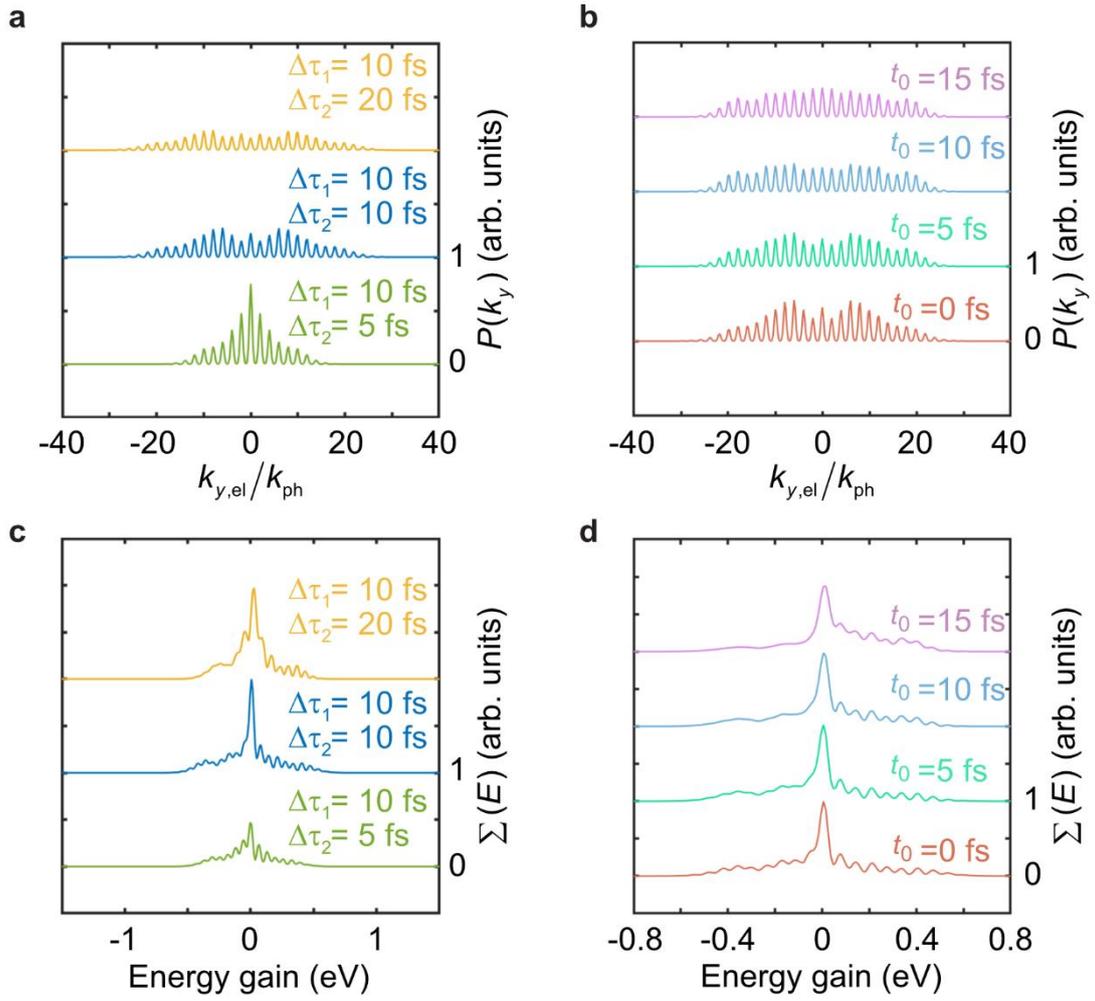

**Figure S10** Final transversal momentum (a, b) and energy-gain (c, d) spectra for the variation of the laser pulse duration $\Delta\tau_1$ and $\Delta\tau_2$ (a, c) and synchronization between electron wavepacket and the arrival time $t_0$ of the laser pulses (b, d). All obtained calculations consider a Gaussian electron wavepacket interaction with two counterpropagating pulsed first-order Hermite-Gaussian (HG$_{10}$) beams (the laser electric-field-amplitude and wavelength for each beam are 4.3 GVm$^{-1}$ and 300 nm, respectively). The calculations presented in (b, d) consider a fixed pulse duration for both counterpropagating beams of $\Delta\tau = 10$ fs. The electron wavepacket has an initial carrier energy of 1 keV and initial longitudinal and transverse FWHM of 250 nm and 60 nm, respectively.

## Supplementary Note 9: Dis-entangling field intensity and pulse duration

The respective influence of the laser pulse duration and field intensity have not been discussed yet in the context of the elastic and inelastic interactions. Thus, we investigate now the interaction of an electron wavepacket with $W_L = 250$ nm, $W_T = 60$ nm and the center kinetic energy of 1 keV, with two counterpropagating $HG_{10}$ pulsed optical beams with central wavelength of $\lambda_1 = 300$ nm and $\lambda_2 = 150$ nm, beam waist of $w_0 = 2\lambda$. We decided to vary both the electric-field amplitudes and pulse durations. The calculations in Figure S10 a-c consider 10 GVm$^{-1}$ and $\Delta\tau = 20$ fs . In Figure S11 d-f both beams are set to 15 GVm$^{-1}$ and $\Delta\tau = 10$ fs. It becomes clear that a longer pulse duration strengthens the transversal momentum interaction even though the field intensity is reduced.

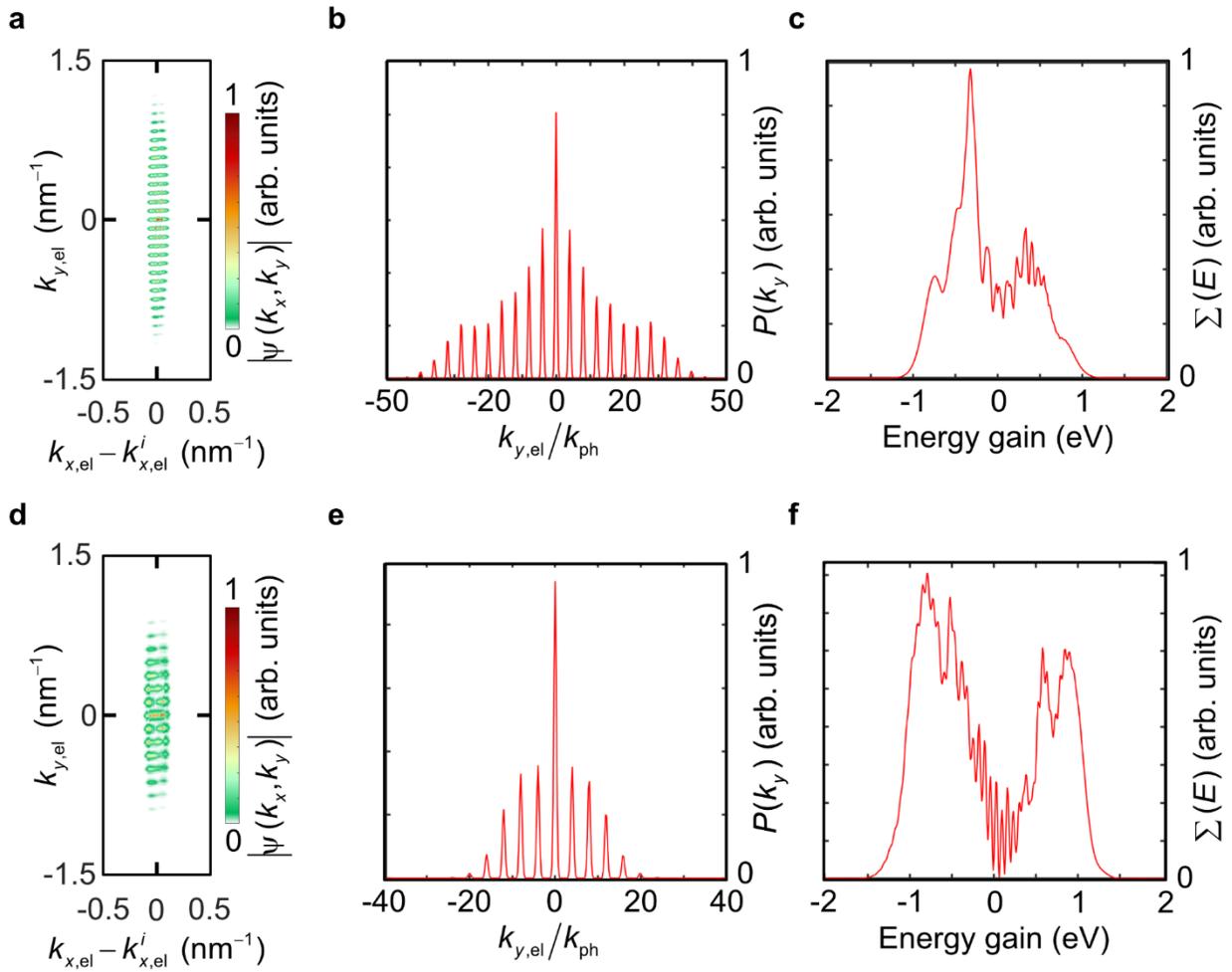

**Figure S11** Elastic and inelastic scattering of an electron wavepacket by a $HG_{10}$ pulsed standing-wave light of varied pulse duration and field intensity. The parameters involved are $W_L = 250$ nm, $W_T = 60$ nm and the center kinetic energy of 1 keV, for the electron wavepacket interacting with two counterpropagating $HG_{10}$ pulsed optical beams with central wavelength of $\lambda_1 = 300$ nm and $\lambda_2 = 150$ nm, and beam waist of $w_0 = 2\lambda$. The calculations in a, b, c consider a field intensity and FWHM pulse duration of 10 GVm$^{-1}$ and $\Delta\tau = 20$ fs respectively. In d, e, f both beams are set to a field intensity and FWHM pulse duration of 15 GVm$^{-1}$ and $\Delta\tau = 10$ fs, respectively. (a, d) Depiction of the transversal and longitudinal momentum distributions for the two different calculations. (b, e) Final transversal momentum distributions normalized to the photon momentum

for the two different calculations. (c, f) Calculated final electron energy gain spectra for both cases stated above.

**Supplementary Note 10: Variation of the electron velocity**

Here, we provide a brief discussion about the influence of the electron velocity on the elastic and inelastic interaction (Figure S12). Therefore, we investigate the interaction of an electron wavepacket with $W_L = 250$ nm, $W_T = 60$ nm and the center kinetic energy of 1 keV, 4 keV and 5 keV, with two counterpropagating $HG_{10}$ pulsed optical beams with central wavelength of $\lambda = 600$ nm, pulse duration of $\Delta\tau = 5$ fs, beam waist of $w_0 = 2\lambda$ and electric-field amplitudes of 4.3 GVm$^{-1}$. It becomes clear that with larger electron velocities the elastic and inelastic interaction strength reduces. Furthermore, it is observable that the sideband separation increases for faster electrons, as expected from the inelastic stimulated Compton scattering process.

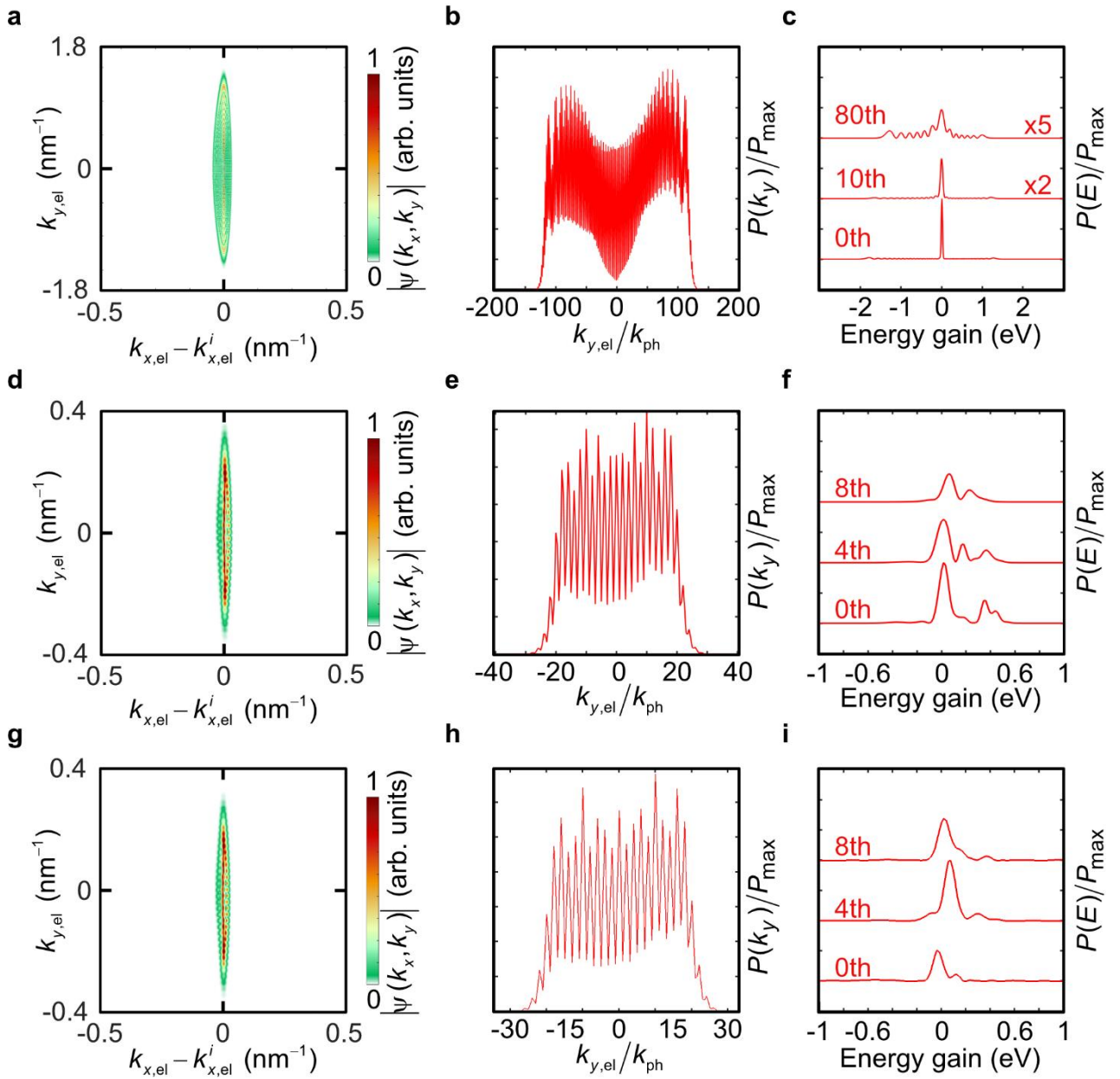

**Figure S12** Elastic and inelastic scattering of an electron wavepacket by pulsed and spatially structured standing wave light beams versus the initial electron kinetic energy of 1 keV (a, b, c), 4 keV (d, e, f) and 5

keV (g, h, i). The pulses have a FWHM temporal durations of 5 fs and the standing-wave light fields have a $HG_{10}$ amplitude structure. Results for an electron wavepacket has an initial kinetic energy of 1 keV. a) Calculated transversal and longitudinal momentum distributions for all populated final electron momentum states. b) Close-up of the final transversal and longitudinal momentum population. c) Final transversal momentum distributions in orders of the photon momentum after the interaction. d) Final electron energy-gain spectra of selected transverse momentum orders. The considered light fields have the electric-field-amplitude and wavelength of 4.3 GVm$^{-1}$ and 600 nm, respectively. The electron wavepacket an initial longitudinal and transverse FWHM broadening of $W_\mathrm{L} = 250$ nm and $W_\mathrm{T} = 60$ nm, respectively.

**Supplementary note 11: Interaction dynamics for different spatial modes**

Here, we discuss the dynamics of the interactions simulated for the interaction of a Gaussian electron wavepacket with $W_L = 250$ nm, $W_T = 60$ nm and the center kinetic energy of 1 keV with two counterpropagating beams of varied spatial mode or wavelength (we discuss the scenario of $HG_{10}$ and $HG_{00}$, respectively). We consider beams of central wavelength of $\lambda = 300$ nm and $\lambda = 150$ nm, pulse duration of $\Delta\tau = 10$ fs, beam waist of $w_0 = 2\lambda$ and electric field amplitude of 5 GVm$^{-1}$ ($HG_{00}$) and 4.3 GVm$^{-1}$ ($HG_{10}$). The dynamics obtained for the interaction with the mixed $HG_{10}$ and $HG_{00}$ spatial modes are depicted in figure S13 a, b. The dynamics obtained for the $HG_{10}$ spatial modes of different wavelength are shown in figure S13 c, d.

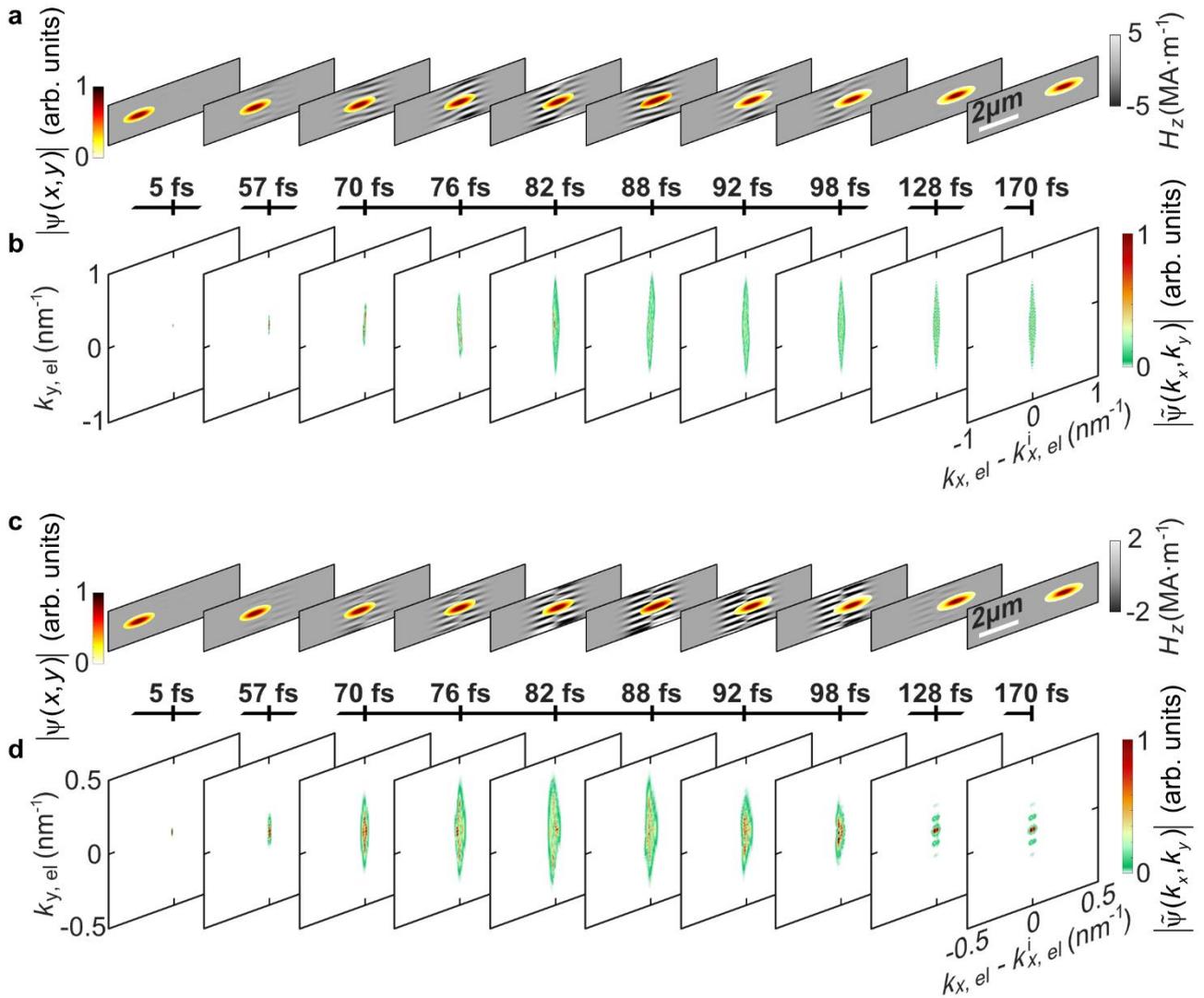

**Figure S13** Dynamics of the evolution of a Gaussian electron wavepacket in the spatial and momentum spaces through two counterpropagating beams of varying spatial modes and wavelengths at different selected time steps. (a, b) $HG_{00}$ (300nm) and $HG_{10}$ (300nm); (c, d) $HG_{10}$ (300nm) and $HG_{10}$ (150nm). The electric-field-amplitude, wavelength, and temporal FWHM broadenings are given by 5 GVm$^{-1}$ ($HG_{00}$) and 4.3 GVm$^{-1}$ ($HG_{10}$), 300 nm and 10 fs, respectively. The electron wavepacket has an initial carrier energy of 1 keV and initial longitudinal and transverse FWHM of 250 nm and 60 nm, respectively. (a, c) The z-component of the magnetic field representing the formed structured-light field (gray background) at depicted time steps, with the insets

demonstrating the amplitude of the electron wavepacket. (b, d) Electron momentum distribution at the corresponding time steps.

## Supplementary References